\documentclass[article,notitlepage,superscriptaddress,twocolumn,%
prd,longbibliography]{revtex4-1}
\usepackage{amssymb}
\usepackage{graphicx}
\usepackage{bm}
\usepackage{bbm}
\usepackage{bbm}
\usepackage{maybemath}
\usepackage{hyperref}
\usepackage{comment}
\usepackage{siunitx}
\usepackage{cancel}
\usepackage{bm}
\usepackage{bbm}
\usepackage{fancyvrb}
\usepackage{xcolor}
\usepackage[NoQandA]{optional}

\newcommand{\dd}{\mathrm{d}}
\newcommand{\ii}{\mathrm{i}}
\newcommand{\ee}{\mathrm{e}}

\def\calG{{\mathcal{G}}}
\def\calP{{\mathcal{P}}}

\newcommand{\app}{\left(\dfrac{\alpha}{\pi}\right)}

\newcommand{\rmr}{{\rm (r)}}

\newcommand{\rmR}{{\rm R}}

\newcommand{\Li}{{\rm Li}}

\newcommand{\Tr}{{\rm Tr}}
\newcommand{\oo}{{\infty}}

\newcommand{\calO}{{\mathcal{O}}}

\newcommand{\FORM}{{\tt FORM}}

\definecolor{garrosgreen}{rgb}{0.1, 0.4, 0.1}
\definecolor{dartmouthgreen}{rgb}{0.05, 0.5, 0.06}
\definecolor{duelferred}{rgb}{0.7, 0.2, 0.1}
\definecolor{cambridgeblue}{rgb}{0.1, 0.3, 1.0}
\definecolor{oxfordblue}{rgb}{0.05, 0.2, 0.7}

\newcommand{\dimmu}{\mbox{\sl\textmu}}

\allowdisplaybreaks

\newcommand{\addrPadova}{Dipartimento di Fisica e Astronomia,
Universit\`a di Padova e 
Istituto Nazionale di Fisica Nucleare, 
Sezione di Padova, Via Marzolo 8, I-35131 Padova, Italy}

\newcommand{\addrRolla}{Department of Physics and LAMOR,
Missouri University of Science and Technology,
Rolla, Missouri 65409, USA}

\begin{document}

\title{Dimensional Regularization and Two--Loop 
Vacuum Polarization Operator: \\
Master Integrals, Analytic Results and Energy Shifts}

\author{Stefano Laporta}
\affiliation{\addrPadova}

\author{Ulrich D. Jentschura}
\affiliation{\addrRolla}

\begin{abstract}
We present a complete reevaluation of the 
irreducible two-loop vacuum-polarization 
correction to the photon propagator in 
quantum electrodynamics, i.e. with an electron-positron
pair in the fermion propagators. 
The integration is carried out by reducing the 
integrations to a limited set of master integrals,
which are calculated using integration-by-parts identities.
Dimensional regularization is used in
$D = 4 - 2\epsilon$ dimensions, and on-mass shell
renormalization is employed.
The one-loop effect is given to 
order $\epsilon$, to be combined with the 
$1/\epsilon$ divergence of the two-loop amplitude.
Master integrals are given.
Final evaluations of two-loop energy shifts for $1S$, $2S$, and $2P$
states are done analytically,
and results are presented, with an emphasis on muonic hydrogen.
For relativistic Dirac--Coulomb reference states,
higher-order coefficients are obtained
for the $Z\alpha$-expansion.
We compare the results obtained to 
the existing literature.
\end{abstract}

\maketitle

\tableofcontents

%
%
\section{Introduction}
\label{sec1}

The two-loop vacuum-polarization correction 
to bound-state energy levels
is an important contribution to the 
Lamb shift in muonic bound systems. 
Specifically, the irreducible two-loop effect
lowers the 2S level energy in muonic hydrogen 
by roughly 1.25~meV in comparison to the 
$2P$ level~\cite{Pa1996mu,Je2011aop1}. 
This energy shift needs to be compared to the
proton-size discrepancy commonly referred to 
as the proton radius puzzle (0.88 fm versus 0.84 fm, 
for a recent brief summary see Ref.~\cite{Je2022proton}).
The proton radius puzzle 
corresponds to a 0.3~meV shift of the $2S$ states,
when expressed in energy units~\cite{Pa1996mu,Je2011aop1}.
The two-loop effect is thus phenomenologically
extremely relevant; any conceivable inaccuracy in its
evaluation could lead to at least a partial theoretical
explanation of the proton size discrepancy.
Vacuum-polarization effects are drastically
enhanced in muonic as compared to 
electronic bound systems~\cite{Pu1957,JeSoIvKa1997,KaIvJeSo1998,KaJeIvSo1998}.
Here, we present a reevaluation of the
irreducible two-loop diagrams, on the basis of modern
integration-by-parts techniques,
employing dimensional regularization.
We note that dimensional regularization
was not yet sufficiently developed in 1973
(despite Refs.~\cite{tHVe1972,BoGi1972}
which appeared in the literature at the time)
to make an evaluation of the two-loop effect 
using dimensional regularization
techniques feasible~\cite{BaRe1973}.
For vacuum-polarization effects in particular, the
modern techniques
lead to drastic simplifications of the calculations.
We find both the imaginary as well as the real part
of the two-loop vacuum tensor in closed analytic form. 

We also derive analytic expressions for
the expectation value of the two-loop potential,
evaluated on (nonrelativistic)
Schr\"{o}dinger--Coulomb eigenstates,
generalizing the treatment originally 
outlined in Ref.~\cite{Pu1957} 
to two-loop order.
In general, for predominantly nonrelativistic 
bound systems, energy shifts can be represented 
in terms of a semi-analytic expansion (in powers 
of $Z\alpha$ and $\ln[(Z\alpha)^{-2}]$, 
where $\alpha$ is the fine-structure constant 
and $Z$ is the nuclear charge number).
As a byproduct, we derive a few higher-order coefficients
from the expectation value of two-loop
potential, evaluated on the relativistic 
Dirac--Coulomb eigenstates.

This paper is organized as follows.
We start by briefly discussing the evaluation of the 
one-loop vacuum-polarization effect in Sec.~\ref{sec2}.
Throughout this paper,
we use dimensional regularization with the 
number $D$ of dimensions expressed as
\begin{equation}
D = 4-2\epsilon \,.
\end{equation}
The one-loop effect needs to be evaluated to order $\epsilon$,
because finite contributions are generated at two-loop order
when the one-loop terms of order $\epsilon$ are multiplied 
by the $1/\epsilon$-terms from the two-loop amplitudes.
The irreducible two-loop diagrams are  discussed in Sec.~\ref{sec3},
including master integrals and their series expansions,
as well as complete results.
A comparison to the available literature is also performed.
In Sec.~\ref{sec4}, we demonstrate that 
energy shifts due to the two-loop effect can be evaluated 
analytically, for muonic bound systems.
Conclusions are reserved for Sec.~\ref{sec5}.

%
%
\section{One--Loop Vacuum Polarization}
\label{sec2}

%
%
\subsection{Orientation}

Our aim is to evaluate the one-loop vacuum-polarization
effect to order $\epsilon$, i.e.,
in a form suitable for the later two-loop calculation.
For simplicity we set the lepton mass
(electron mass) in the vacuum-polarization loops 
to be  $m=1$.
We use for a metric tensor 
$g_{\mu\nu} = \text{diag}(1,-1,-1,-1)$
(see, for example, Refs.~\cite{BjDr1965,PeSc1995}),
where 
\begin{align}
p^{\mu}=& \; (p_0,\vec{p}), 
\qquad
p_{\mu}=(p_0,-\vec{p}), 
\qquad
\\[0.1133ex]
\{\gamma^{\mu},\gamma^{\nu}\}=& \; 2g^{\mu\nu},
\qquad
S(p)= \frac{1}{\cancel{p} - m + \ii \epsilon} \,,
\qquad 
m=1 \,.
\end{align}
Here, $m=1$ is the electron mass.

Due to current conservation,
the vacuum polarization function $\Pi_{\mu\nu}$ has this tensorial
structure 
\begin{equation}
\Pi_{\mu\nu}(q^2)=
( q_{\mu}q_{\nu} -q^2 \, g_{\mu\nu}) \, \Pi(q^2)\,.
\end{equation}
We follow the convention of Ref.~\cite{ItZu1980} so that
\begin{equation}
\label{derivative}
\left. \frac{\dd\Pi(q^2)}{\dd q^2}
\right|_{q^2=0} > 0 
\end{equation}
at the one-loop and two-loop level.
For completeness, we note that the 
opposite conventions are used elsewhere 
in the literature, e.g., in 
Refs.~\cite{BeLiPi1982vol4,JeAd2022book}.
The conventions employed here are
consistent with 
those employed in Ref.~\cite{ItZu1980}.
Summation over the Lorentz indices in $D$ dimensions leads
to the result
\begin{equation}
{\Pi^\mu}_\mu(q^2)=(-q^2 \, {g^\mu}_\mu + q^{\mu}\, q_{\mu}) \, \Pi(q^2)
= q^2 \, (1-D) \, \Pi(q^2)\,.
\end{equation}
So, we can extract the scalar vacuum polarization
function $\Pi(q^2)$ as
\begin{equation}
\Pi(q^2)=\frac{ {\Pi^\mu}_\mu(q^2)}{q^2 \, (1-D)}\,.
\end{equation}
The threshold for pair production of 
$\Pi(q^2=s)$ is $q^2 = s = 4$.
The argument of the polarization tensor is 
\begin{equation}
q^2 = (q^0)^2 - \vec q^{\,2} \,,
\end{equation}
which is the four-momentum square of the photon
entering the vacuum-polarization loop.

%
%
\subsection{Calculation}

For the renormalization of the two-loop diagram with a self-mass insertion
we need the one-loop vacuum polarization function $\Pi^{(1)}$,  
and the function $\Pi^{(1a)}$ from the same diagram with 
iterated electron propagators on one side of the loop.
The unrenormalized tensors read as follows,
\begin{align}
\Pi^{(1)}_{\mu \nu}(q^2) =& \;
{ -\ii}
e^2 \int \dfrac{\dd^D k}{(2\pi)^D} 
\left[ -\Tr\left(
\gamma_{\mu}S(k)\gamma_\nu S(k - q) \right) \right]\,,
\\
\Pi^{(1a)}_{\mu \nu}(q^2) =& \;
{ \ii}
e^2 \int \dfrac{\dd^D k}{(2\pi)^D}
\left[ -\Tr\left(
\gamma_{\mu} [S(k)]^2 \gamma_\nu S(k - q) \right) \right]\,.
\end{align}
The needed quantities are expressed in terms of two one-loop master
integrals,
\begin{subequations}
\begin{align}
\label{Pi1}
\Pi^{(1)}(q^2) =& \; 
\left( \frac{e^2 \, C(\epsilon)}{4 \pi \alpha} \right) \,
\app \, P^{(1)}(q^2) \,,
\\[0.1133ex]
\label{Pi1a}
\Pi^{(1a)}(q^2) =& \; 
\left( \frac{e^2 \, C(\epsilon)}{4 \pi \alpha} \right) \,
\app \, P^{(1a)}(q^2) \,.
\end{align}
\end{subequations}
The prefactor in these results simplifies to unity 
provided we set
\begin{equation} 
\label{id}
e^2 = \frac{ 4 \pi \alpha }{ C(\epsilon) } \,,
\qquad
C(\epsilon) =
\Gamma(1+\epsilon) \; (4\pi)^{\epsilon} \; {\dimmu}^{-2\epsilon} \,.
\end{equation}
One observes that the identification~\eqref{id}
is not completely canonical 
in dimensional regularization.
Namely [see, e.g., Eq.~(10.173) of Ref.~\cite{JeAd2022book}],
in the $\overline{\mathrm{MS}}$ scheme, one normally
has the relation
$e^2 = (4 \pi)^{1 - \epsilon} \, \alpha \,
\dimmu^{2\epsilon} \ee^{\gamma_E \epsilon}$,
where $\gamma_E = 0.57721\,56649\dots$
is the Euler--Mascheroni constant,
and $\dimmu$ is the renormalization scale
induced by dimensional regularization.
This is equivalent to Eq.~\eqref{id} 
up to terms linear in $\epsilon$.
The prefactor $1/C(\epsilon)$ helps to 
simplify results after integration.

The one-loop integrals are as follows,
\begin{align}
\label{redu1}
P^{(1)}(q^2) =& \;
\frac{(D-2) q^2+4}{2 (D-1) q^2}
M_{12}(q^2) +
\frac{(D-2) M_{2}(q^2)}{(D-1) q^2} \,,
\\
P^{(1a)}(q^2) =& \;
\frac{(D-4) q^2+4}{q^2 \left(q^2-4\right)}
M_{12}(q^2) 
+ \frac{2 (D-2) M_{2}(q^2)}{q^2 \left(q^2-4\right)} \,,
\end{align}
where 
\begin{subequations}
\begin{align}
M_{12}(q^2)=& \; N(\epsilon) \int \frac{[\dd^D k_1] }{D_1 D_2} \,,
\\[0.1133ex]
M_{2}(q^2) =& \; N(\epsilon) \int \frac{[\dd^D k_1] }{D_2} \,,
\\[0.1133ex]
D_j =& \; -p_j^2+1- \ii \epsilon\,, \qquad j=1,\ldots,2\,.
\\[0.1133ex]
p_1 =& \; k_1 \,,
\qquad
p_2 = k_1 - q \;\; \,,
\\[0.1133ex]
[\dd^D k]=& \dimmu^{-2\epsilon} \dd^D k \,.
\end{align}
\end{subequations}
The above definition of $[\dd^D k]$ 
contains the factor $\dimmu^{-2\epsilon}$.
It simplifies with the factor
$1/\dimmu^{-2\epsilon}$ of $N(\epsilon)$ 
so that the master integrals are independent of $\dimmu$.
The normalization factor is
\begin{equation}
\label{defN}
N(\epsilon) 
= \frac{\dimmu^{2\epsilon} }{\pi^{2 - \epsilon} \; \Gamma(1+\epsilon) }
= \frac{4^{\epsilon}}{\pi^{2-2\epsilon} C(\epsilon)}
= \frac{4(4\pi^2)}{(2\pi){^{4-2\epsilon}} C(\epsilon)}\,.
\end{equation}
The latter form exhibits the dimension $D = 4-2\epsilon$.
With the normalization factor $N(\epsilon)$, all master
integrals $M$ occurring in our calculations are
dimensionless and not only uncluttered from logarithms of $\dimmu$,
but also, from logarithms of $4\pi$.

One can show that the 
one-loop master integrals satisfy the following 
differential equations,
\begin{subequations}
\begin{align}
& M_{2}'\left(q^2\right) =0 \,,
\\[1.1133ex]
& 
\left[(D-4) q^2+4\right]
\, M_{12}(q^2) + 2 (D-2) M_{2}\left(q^2\right) 
\nonumber\\
&
-2 \left(q^2-4\right) q^2 
M_{12}'\left(q^2\right) =0 \ ,
\end{align}
\end{subequations}
where the differentiation is with respect to $q^2$.
One finds that $M_2$ is a momentum-independent 
vacuum integral, 
\begin{equation}\label{m2}
M_2(q^2)=-\frac{1}{\epsilon(1-\epsilon)}  \,.
\end{equation}
For $M_{12}(q^2)$, one writes the result 
as a function of the variable $v = v(q^2)$,
\begin{equation}
\label{vdef}
v = \sqrt{1- \frac{4}{q^2}} \,.
\end{equation}
The result is 
\begin{align}
\label{m12}
(1 - & 2\epsilon) M_{12}(q^2)=\frac{1}{\epsilon}
+ v \ln \left(\frac{v-1}{v+1}\right)
-\frac{\epsilon}{6} v \biggl[ \pi^2 + 
\nonumber\\ 
& 12 \text{Li}_2\left(\frac{1-v}{v+1}\right)
+ 12 \ln \left(\frac{2 v}{v+1}\right)
\ln \left(\frac{v-1}{v+1}\right)
\nonumber\\ 
& -3 \ln^2\left(\frac{v-1}{v+1}\right) \biggr]
+ \epsilon^2 \frac{v}{6} N_{12}(v)
+ \calO(\epsilon^3) \,,
\end{align}
where we include the terms of order $\epsilon$ and $\epsilon^2$
which will be useful in the context of the two-loop 
calculation. The explicit form of $N_{12}(q^2)$ is 
given in Appendix~\ref{appa}.

%
%
\subsection{Renormalization}

Substituting Eqs.~\eqref{m2} and~\eqref{m12} into \eqref{redu1},
we can find the expansions of $P^{(1)}(q^2)$ and $P^{(1a)}(q^2)$.
But for the renormalization we need to subtract the values 
at $q^2=0$,
\begin{equation}
P^{(1)}(0) = \frac{1}{3 \epsilon}\ ,
\qquad
P^{(1a)}(0) = \frac{1}{3}\, .
\end{equation}
In our formalism, the scaled 
functions $P^{(1)}$ and $P^{(2)}$ do not 
have powers of $\dimmu$, but the 
scalar vacuum polarization
functions $\Pi^{(1)}$, $\Pi^{(2)}$ depend on $\dimmu$.
Yet, after renormalization the dependence 
disappears in the limit $\epsilon \to 0$.
The renormalized expression for the one-loop
function $P^{(1)}_{\rm R}(q^2)$ is as follows,
\begin{multline}
\label{P1R}
P^{(1)}_{\rm R}(q^2) = P^{(1)}(q^2) - P^{(1)}(0)
\\
= \frac{1}{9} \left(8-3 v^2\right) -
\frac{v}{6} \left(v^2-3\right) \ln \left(\frac{v-1}{v+1}\right)
\\
+ \epsilon \, Q^{(1)}_1(v) 
+ \epsilon^2 \, Q^{(1)}_2(v) + \calO(\epsilon^3) \,.
\end{multline}
The coefficient of the term linear in $\epsilon$ is 
\begin{multline}
\label{P1Rlinear}
Q^{(1)}_1(v) = 
\frac{v}{36} \left(v^2-3\right) 
\biggl[12 \text{Li}_2\left(\frac{1-v}{v+1}\right)
\\ 
-3 \ln^2\left(\frac{v-1}{v+1}\right) +
12 \ln \left(\frac{2 v}{v+1}\right) \,
\ln \left(\frac{v-1}{v+1}\right) +\pi ^2 \biggr]
\\
+ \left(v-\frac{4 v^3}{9}\right) 
\ln\left(\frac{v-1}{v+1}\right)-\frac{8 v^2}{9}+\frac{52}{27} \,.
\end{multline}
For the explicit form of the term of order $\epsilon^2$,
we refer to Appendix~\ref{appa}.

The renormalized value for the function
$P^{(1a)}_{\rm R}(q^2)$ with iterated electron propagators 
is as follows,
\begin{multline}
\label{P1aR}
P^{(1a)}_{\rm R}(q^2) = 
P^{(1a)}(q^2)-P^{(1a)}(0)
\\ 
= \frac{v^2-1}{4v} \biggl\{
\frac{2 v \left(3 v^2-5\right)}{3 \left(v^2-1\right)}+\left(v^2-1\right) \ln
\left(\frac{v-1}{v+1}\right)
\biggr\}
\\
+ \epsilon Q^{(1a)}_1(v) 
+ \epsilon^2  Q^{(1a)}_2(v) 
+ \calO(\epsilon^3) \,.
\end{multline}
The term linear in $\epsilon$ is given as follows,
\begin{multline}
\label{P1aRlinear}
Q^{(1a)}_1(v) = 
\frac{v^2-1}{4v} \biggl[
\left(v^2-1\right)
\biggl( -2 \text{Li}_2\left(\frac{1-v}{v+1}\right)
\\ 
+ \frac{1}{2} \ln^2\left(\frac{v-1}{v+1}\right)
- \frac{\pi^2}{6} \biggr)
+ 2 \ln\left( \frac{v-1}{v+1} \right)
\\
\times \biggl( v^2-\left(v^2-1\right) 
\ln\left(\frac{2 v}{v+1}\right)
\biggr) +4 v \biggr]
\end{multline}
The quadratic term in $\epsilon$, denoted as $Q^{(1a)}_2(v)$,
is given in Appendix~\ref{appa}.
Actually, only the terms proportional to $\epsilon$ are necessary for the
renormalization of the two-loop vacuum polarization,
but higher-order terms are useful for higher-loop
calculations, and are thus relegated to Appendix~\ref{appa}. 
 
In the limit $\epsilon \to 0^+$, one recovers the 
known result for the one-loop vacuum polarization,
\begin{multline}
\label{Pi1R}
\Pi^{(1)}_\rmR(q^2) = \app P^{(1)}_\rmR(q^2)
\\ 
= \app\left[
\frac{1}{9} \left(8-3 v^2\right)-\frac{1}{6} v \left(v^2-3\right) \ln
\left(\frac{v-1}{v+1}\right) \right] \,.
\end{multline}
The dispersion relation is
\begin{equation}
\Pi^{(1)}_{\rm R}(q^2) = 
\frac{q^2}{\pi}
\int_{4}^\infty \dd q'^2
\frac{{\rm Im} \, 
\Pi^{(1)}_{\rm R}(q'^2 + \ii \epsilon)}{q'^2 ( q'^2 - q^2 )} \,.
\end{equation}
The imaginary part is positive
infinitesimally above the cut, 
and negative infinitesimally below the cut.
One has the simple representation
\begin{equation}
\label{ImPi1R}
{\rm Im} \, \Pi^{(1)}_{\rm R}(q^2 + \ii \epsilon)
= \frac{\alpha}{6} v (3 - v^2) \, \Theta(q^2 - 4) \,.
\end{equation}
On the cut, the $v$ variable runs from $v=0$ 
to $v=1$.
Here, $\Theta$ denotes the Heaviside step function.
The above result 
given in Eq.~\eqref{Pi1R} in the 
$v$ representation fulfills
[see also Eq.~\eqref{derivative}]
\begin{equation}
\label{Pi1Rlow}
\Pi^{(1)}_{\rm R}(q^2) 
=  \frac{\alpha}{\pi} \frac{q^2}{15} + \calO(q^4) \,,
\end{equation}
for $q^2 = (q^0)^2 - \vec q^{\,2}$.

%
%
\section{Two--Loop Vacuum Polarization}
\label{sec3}

%
%
\subsection{Orientation}

\begin{figure}[t!]
\begin{center}
\begin{minipage}{0.99\linewidth}
\begin{center}
\includegraphics[width=0.99\linewidth]{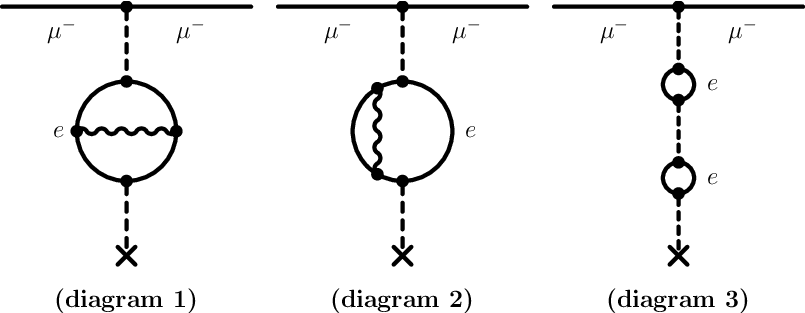}
\end{center}
\caption{\label{fig1} The diagrams concern the 
electronic two-loop vacuum-polarization diagrams in muonic hydrogen
(the negatively charged muon line is denoted by $\mu^-$).
These are are naturally divided into diagram~(1), which
is a proper two-loop diagram,
diagram~(2), which is a one-loop diagram
with a self-energy insertion,
and the loop-after-loop (reducible) diagram~(3).}
\end{minipage}
\end{center}
\end{figure}

We focus on the diagrams in Figs.~\ref{fig1}.
The first diagram~1 is the proper two-loop vacuum polarization diagram,
while diagram~2 is a one-loop vacuum polarization diagram with a
self-energy insertion.
The expression $\Pi^{(2;1)}_{\mu \nu}(q^2)$ 
corresponds to diagram~1,
\begin{subequations}
\label{spinol}
\begin{align}
& \Pi^{(2;1)}_{\mu \nu}(q^2) = 
e^4 \int \dfrac{\dd^D k_1}{(2\pi)^D} \dfrac{\dd^D k_2}{(2\pi)^D} 
\nonumber\\
& \; \times \left[ -\Tr\left(
\gamma_{\mu}S(p_4) \, \gamma^\rho \, S(p_2)\gamma_{\nu}
S(-p_1)\gamma^\sigma \, S(-p_3) \, D_{\rho \sigma}(p_5)
\right)
\right] \,,
\end{align}
while $\Pi^{(2;2)}_{\mu \nu}(q^2)$ 
corresponds to the self-energy insertion
into a fermionic line of the vacuum-polarization loop,
\begin{align}
& \Pi^{(2;2)}_{\mu \nu}(q^2) = 
e^4 \int \dfrac{\dd^D k_1}{(2\pi)^D} \dfrac{\dd^D k_2}{(2\pi)^D} 
\nonumber\\
& \; \left[ -\Tr\left(
\gamma_{\mu} \, S(p_4) \, \gamma^\rho \, S(p_2) \,
\gamma^\sigma \, S(p_4) \, \gamma_\nu \, S(-p_3) \,
D_{\rho \sigma}(p_5) \right) \right]\,.
\end{align}
\end{subequations}
Here, the photon propagator 
is given by [see Eq.~(9.133) of Ref.~\cite{JeAd2022book}]
\begin{equation}
\label{photonprop}
D_{\rho\sigma}(k) = 
- \frac{1}{k^2 + \ii \epsilon}
\left( g_{\rho\sigma} -
\lambda \frac{ k_{\rho} k_{\sigma}}{k^2} \right) \,,
\end{equation}
and the momenta and denominators are 
\begin{subequations}
\label{andthen}
\begin{align}
p_1= & \; k_1\,, \qquad p_2 = q-k_1\,, 
\qquad p_3= k_2\,,
\\[0.1133ex]
p_4 =& \; q-k_2\,, \qquad
p_5 = k_1-k_2 \,,
\\[0.1133ex]
D_j =& \; -p_j^2+1- \ii \epsilon\,, \qquad j=1,\ldots,4\,, 
\\[0.1133ex]
D_5 =& \; -p_5^2 - \ii \epsilon \,.
\end{align}
\end{subequations}
These conventions are adapted to 
\FORM~\cite{FORM,KuUeVeVo2013}; in an East--Coast metric
with $g_{\mu\nu} =  \mathrm{diag}(-1,-1,-1,1)$, 
one would replace $p_j^2 \to -p_j^2$.

%
%
\subsection{Reduction to Master Integrals}

After the calculation of traces, performed with computer algebra
(using \FORM, see Refs.~\cite{FORM,KuUeVeVo2013}), and a Wick rotation, the scalar
contribution $\Pi^{(1)}(q^2)$ and $\Pi^{(2)}(q^2)$ of the two diagrams are
reduced to a combination of 24 and 41 Feynman integrals, respectively.  In
order to further reduce the expressions, we use the integration by parts (IBP)
identities~\cite{ChTk1981npb,Tk1981plb}. A system of IBP identities is built
and solved with the program $\tt{SYS}$,
which is based on an algorithm described in 
Refs.~\cite{La2000ijmpa,La2001dc}.
These identities allow to reduce $\Pi^{(1)}(q^2)$ and
$\Pi^{(2)}(q^2)$ to a linear combination of five irreducible master integrals.

The choice of the master integrals is somewhat arbitrary; 
we chose  the following ones: 
\begin{subequations}
\begin{align}
M_{1234} = & \; [N(\epsilon)]^2 
  \int \frac{[\dd^D k_1] [\dd^D k_2]}{D_1 D_2 D_3 D_4} \,,
\\[0.1133ex]
M_{234}  = & \; [N(\epsilon)]^2 
  \int \frac{[\dd^D k_1] [\dd^D k_2]}{    D_2 D_3 D_4} \,,
\\[0.1133ex]
M_{235} = & \; [N(\epsilon)]^2 
  \int \frac{[\dd^D k_1] [\dd^D k_2]}{    D_2 D_3 D_5} \,,
\\[0.1133ex]
M_{235k} = & \; [N(\epsilon)]^2 
  \int \frac{[\dd^D k_1] [\dd^D k_2] \; q \cdot k_2}{D_2 D_3 D_5}\,,
\\[0.1133ex]
M_{24} =& \; [N(\epsilon)]^2 \int \frac{[\dd^D k_1] [\dd^D k_2]}{D_2 D_4} \,.
\end{align}
\end{subequations}
The following master integrals are reducible 
and factorize into a product of two one-loop
master integrals,
\begin{subequations}
\label{fact}
\begin{align}
\label{m24}
M_{24}(q^2) =& \; [M_2(q^2)]^2\,,
\\
\label{m234}
M_{234}(q^2) =& \; M_2(q^2) M_{12}(q^2) \,,
\\
\label{m1234}
M_{1234}(q^2) =& \; [M_{12}(q^2)]^2 \,.
\end{align}
\end{subequations}
The expressions for the contribution of diagrams~1 and~2 are
given as follows,
\begin{subequations}
\begin{align}
\Pi^{(2;1)}(q^2) =& \;  [C(\epsilon)]^2 \app^2 
  \left(P^{(2;1)}(q^2)+\lambda G^{(2;1)}(q^2)\right) \,,
\\[0.1133ex]
\Pi^{(2;2)}(q^2) =& \;  [C(\epsilon)]^2 \app^2 
  \left(P^{(2;2)}(q^2)+\lambda G^{(2;2)}(q^2)\right) \,,
\end{align}
\end{subequations}
where $\lambda$ is the gauge parameter [see Eq.~\eqref{photonprop}],
and
\begin{multline}
\label{redu21}
P^{(2;1)}(q^2)=
\calP^{(2;1)}_{24} \, M_{24}(q^2) 
+ \calP^{(2;1)}_{234} \, M_{234}(q^2) 
\\
+ \calP^{(2;1)}_{1234} \, M_{1234}(q^2) 
+ \calP^{(2;1)}_{235} \, M_{235}(q^2) 
\\
+ \calP^{(2;1)}_{235k} \, M_{235k}(q^2) \,.
\end{multline}
The coefficient functions
$\calP^{(2;1)}_{24}$,
$\calP^{(2;1)}_{234}$,
$\calP^{(2;1)}_{1234}$, 
$\calP^{(2;1)}_{235}$
and $\calP^{(2;1)}_{235k}$
are given in Appendix~\ref{appb}.
The gauge-dependent part of diagram 1 is given as
\begin{multline}\label{redugauge}
G^{(2;1)}(q^2) =
\calG^{(2;1)}_{24} \, M_{24}(q^2) 
+ \calG^{(2;1)}_{235} \, M_{235}(q^2) 
\\
+ \calG^{(2;1)}_{235k} \, M_{235k}(q^2) \,,
\end{multline}
with the coefficient functions
$\calG^{(2;1)}_{24}$, $\calG^{(2;1)}_{235}$
and $\calG^{(2;1)}_{235k}$
also being indicated in Appendix~\ref{appb}.
For diagram 2, one has the following reduction,
\begin{multline}
\label{redu22}
P^{(2;2)}(q^2) =
\calP^{(2;2)}_{234} \, M_{234}(q^2)
+ \calP^{(2;2)}_{235k} \, M_{235k}(q^2)
\\
+ \calP^{(2;2)}_{235} \, M_{235}(q^2)
+ \calP^{(2;2)}_{24} \, M_{24}(q^2) \,.
\end{multline}
Again, the coefficient functions
$\calP^{(2;2)}_{24}$,
$\calP^{(2;2)}_{234}$,
$\calP^{(2;2)}_{1234}$,
$\calP^{(2;2)}_{235}$,
and $\calP^{(2;2)}_{235k}$
are given in Appendix~\ref{appb}.
The gauge-dependent part of the second
diagram cancels
the gauge-dependent part of the first diagram,
in view of the relation $G^{(2;2)}(q^2) = - \tfrac{1}{2} G^{(2;1)}(q^2)$.

%
%
\subsection{Differential Equations and Master Integrals}

A system of differential equations in $q^2$ is being created;
we need to compute the 5 master integrals.
Three of them are reducible into 
products of one-loop integrals,
as outlined in Eq.(\ref{fact}. 
We obtain a system of differential equations in $q^2$
satisfied by the irreducible  master 
integrals~\cite{Ko1991plb,Re1997ny,GeRe2000}.
The first of these equations is of second
order, 
\begin{subequations}
\begin{align}
& q^2 \left[ 
-3 \left((D-4) q^2+4\right)
M_{235}'(q^2)
\right.
\nonumber\\
& \left. 
+2  q^2 \left(q^2-4\right) 
M_{235}''(q^2) \right]
+ (D-2)^2 M_{24}(q^2)
\nonumber\\
& 
+(D-3) \left[D \left(q^2+2\right)-4 \left(q^2+1\right)\right]
\, M_{235}(q^2)
=0 \,,
\end{align}
while the remaining one is a first-order differential
equation,
\begin{align}
&
\left[ 
D \left(q^2+2\right)-q^2-6
\right] \, 
M_{235}(q^2)
+6 (D-2)
\, M_{235k}(q^2) 
\nonumber\\
& 
+ (D-2)
\, M_{24}(q^2) 
+ q^2 \, \left(q^2-4\right)\,
M_{235}'(q^2)=0 \,, 
\end{align}
\end{subequations}
where the differentiation is with respect to $q^2$.
The initial conditions at $q^2=0$ for these diagrams are simple vacuum
integrals.

%
%
\subsection{Results for the Master Integrals}

Due to the appearance of the factor $D-4$ in the denominators of
Eqs.~\eqref{redu21}---\eqref{redu22} (for the coefficients,
see Appendix~\ref{appb}),
the master integrals have to be evaluated up to order $\calO(\epsilon)$.
It is not sufficient to keep only the finite part.
The results are as follows.
For $M_{235}$, one finds
the following relation, where we note the prefactor $(1-v^2)$,
\begin{align}
\label{m235}
(1 & -v^2) M_{235}(q^2) =\frac{v^2-1}{\epsilon^2}
+\frac{3 v^2-2}{\epsilon}
+\biggl[
  6 v^2
   +\frac{1}{2}
\nonumber\\ &
+\frac{1}{4}
   \left(-v^4-2 v^2+3\right) \ln
   ^2\left(\frac{v-1}{v+1}\right)
\nonumber\\ &
-\left(v^2-3\right) v \ln \left(\frac{v-1}{v+1}\right)
\biggr]
+\epsilon N_{235}(q^2)
+ \calO\left(\epsilon^2\right)\,.
\end{align}
The order-$\epsilon$
coefficient $N_{235}(q^2)$ is found in Appendix~\ref{appc}.
For $M_{235k}$, one finds the relation
\begin{align}
\label{m235k}
(1& -v^2)^2 M_{235k}(q^2) =\frac{1-v^2}{\epsilon^2}
+\frac{1}{\epsilon}\biggl[
\frac{13}{6}-\frac{7 v^2}{2}
\biggr]
\nonumber\\ & 
   +\frac{1}{8}
   \left(-v^6+3 v^4+5 v^2-7\right) \ln ^2\left(\frac{v-1}{v+1}\right)
\nonumber\\ &
  -\frac{1}{6} v \left(3 v^4-8 v^2+21\right) \ln \left(\frac{v-1}{v+1}\right)
\nonumber\\ &
+\frac{1}{12} \left(-6 v^4-91 v^2-7\right)
+\epsilon N_{235k}(q^2)
+ \calO\left(\epsilon^2\right)\,.
\end{align}
Again, $N_{235k}(q^2)$ can be found in Appendix~\ref{appc}.
Similar master integrals have recently 
been considered in Refs.~\cite{ScTa1993,AnEtAl2007}.

%
%
\subsection{Renormalization}

We now substitute the master integrals~\eqref{m24}, \eqref{m1234}, \eqref{m234}, 
\eqref{m235} and~\eqref{m235k}
into the reduction formulas given in Eqs.~\eqref{redu21}
and~\eqref{redu22}. We can thus find
the series expansions of the vacuum-polarization
functions $\Pi^{(2;1)}(q^2)$ and $\Pi^{(2;2)}(q^2)$.
For the renormalization,
we need the values at $q^2 =0$.
The behavior, for small $q^2$, is the following:
\begin{align}
P^{(2;1)}(q^2)&=
-\frac{12 \epsilon^5+4 \epsilon^4-23 \epsilon^3-14 \epsilon^2-21 \epsilon-6}%
{12 (\epsilon-2) (\epsilon-1) \epsilon^2
(2 \epsilon+1)(2 \epsilon+3)} \cr
& -\frac{1}{q^2}\frac{1}{(1-\epsilon)(1+2\epsilon)(1-2\epsilon)} 
+ \calO(q^2)\ , 
\cr
P^{(2;2)}(0)&=
-\frac{(\epsilon+1) \left(4 \epsilon^4-8 \epsilon^3-8 \epsilon^2+
21 \epsilon+3\right)}{12 (\epsilon-2)
(\epsilon-1) \epsilon^2 (2 \epsilon+1) (2 \epsilon+3)}
\cr
& +\frac{1}{2 q^2}\frac{1}{(1-\epsilon)(1+2\epsilon)(1-2\epsilon)} 
+ \calO(q^2)\ .
\end{align}
Terms containing $1/q^2$ appear, due to the fact that the single
diagrams are not gauge invariant.
Taking the gauge invariant sum 
$P^{(2;1)}(q^2)+2 P^{(2;2)}(q^2)$,
the divergent terms cancel out,
so that, taking the limit $q^2 \to 0$ one gets 
\begin{equation}
P^{(2;1)}(0)+2 P^{(2;2)}(0)
= -\frac{10 \epsilon^2 -7 \epsilon -9}{12 (\epsilon-2) \epsilon (2 \epsilon+1)} \,.
\end{equation}

In order to carry out the renormalization, 
we need the renormalization constants

\begin{equation}
Z_J = 1 + \sum_L Z_J^{(L)}\left( \frac{\alpha}{\pi}\right)^L
\left( \frac{e^2 \, C(\epsilon)}{4 \pi \alpha} \right)^L \,,
\end{equation}
where $J$ stands for the 
vertex renormalization ($J=1$), the 
wave function renormalization ($J=2$),
and the mass renormalization ($J=m$),
and $L$ enumerates the loop order.
We define the one-loop quantities
\begin{equation}
\label{FS0S1}
F =-Z_1^{(1)} \,,
\qquad
S_0=\delta m= -Z_m^{(1)} \,,
\qquad
S_1=-Z_2^{(1)} \,,
\end{equation}
and obtain the results
\begin{multline}
F  =\frac{3-2 \epsilon}{4 \epsilon-8 \epsilon^2} \,, \quad
S_0=\frac{3-2 \epsilon}{4 \epsilon-8 \epsilon^2} \,, \quad
S_1= S_0 \,.
\end{multline}
We note that 
$Z_1^{(1)}=Z_2^{(1)}$ because of the Ward--Takahashi identity.
The identity $Z_2^{(1)}=Z_m^{(1)}$ is a simple coincidence, 
which occurs only at one-loop order in dimensional regularization,
but to all orders in $\epsilon$.
The constants $F$, $S_0$ and $S_1$ in 
Eq.~\eqref{FS0S1} are defined with 
the prefactor $(-1)$ so that the expansion of the 
the self-energy insertion 
into the fermion line 
about the mass shell 
acquires positive terms, for $\cancel{q}\approx m=1$:
\begin{align}
\frac{1}{\cancel{q} - m}
\Sigma(q)
\frac{1}{\cancel{q} - m} =& \;
\frac{1}{\cancel{q} - m}
[ S_0+ (\cancel{q} - m) S_1 ]
\frac{1}{\cancel{q} - m} 
\nonumber\\
=& \; \frac{S_1}{\cancel{q} - m} 
+ \frac{S_0}{(\cancel{q}-m)^2} \,.
\end{align}
Finally, the renormalized contributions of the diagrams are
found as follows,
\begin{align}
P^{(2;1)}_{\rm R}(q^2) =& \; P^{(2;1)}(q^2)-P^{(2;1)}(0) 
- 2 F P^{(1)}_{\rm R}(q^2) \,,
\\
P^{(2;2)}_{\rm R}(q^2) =& \; 
P^{(2;2)}(q^2)-P^{(2;2)}(0) 
\nonumber\\
& 
- (S_1 P^{(1)}_{\rm R}(q^2) + S_0 P^{(1a)}_{\rm R}(q^2)) (-1) \,,
\end{align} 
and the renormalized two-loop 
function $P^{(2)}_{\rm R}(q^2)$ is obtained
as
\begin{equation}
P^{(2)}_{\rm R}(q^2) =
P^{(2;1)}_{\rm R}(q^2)+2 P^{(2;2)}_{\rm R}(q^2) \,.
\end{equation}

%
%
\subsection{Results}

We recall the definition of $v$ in
Eq.~\eqref{vdef}.
In the limit $\epsilon \to 0^+$,
one obtains the two-loop
vacuum-polarization function
\begin{equation}
\Pi^{(2)}_\rmR(q^2) = \app^2 \, P^{(2)}_{\rm R}(q^2) \,.
\end{equation}
We find a compact representation (see also Ref.~\cite{LaJe2024suppl})
\begin{multline}
\label{Pi2R}
\Pi^{(2)}_{\rm R}(q^2) = \app^2 \biggl\{
\frac{ v^4 - 2 v^2  -3  }{12}
\biggl[\Phi_1\left(\frac{v-1}{v+1}\right) 
\\
\times \ln^2\left(\frac{v-1}{v+1}\right) 
- 4 \Phi_2\left(\frac{v-1}{v+1}\right) 
\ln\left(\frac{v-1}{v+1}\right) + 
6 \Phi_3\left(\frac{v-1}{v+1}\right) 
\\
+ 3 \zeta (3)\biggr]  
+\frac{3 v - v^3}{12} \biggl[ 4 \Phi_1\left(\frac{v-1}{v+1}\right) 
\ln\left(\frac{v-1}{v+1}\right) 
\\ 
-4 \Phi_2\left(\frac{v-1}{v+1}\right) 
+ 3 \ln^2\left(\frac{v-1}{v+1}\right) \biggr] 
+ \frac{ 5 v - 3 v^3}{8} \ln \left(\frac{v-1}{v+1}\right) \\
+\frac{1}{96} \left(7 v^4-22 v^2-33\right) 
\ln^2\left(\frac{v-1}{v+1}\right)
- \frac{13 (v^2-1)}{24} +\frac{5}{24}
\biggr\} \,,
\end{multline}
where
\begin{equation}
\label{defPhi}
\Phi_n(x)=\Li_n(x) + 2 \Li_n(-x)\ , \quad \Li_1(x)=-\ln(1-x)\ .
\end{equation}
For small positive $q^2$, one obtains the 
expression
\begin{equation}
\label{Pi2Rlow}
\Pi^{(2)}_\rmR(q^2) = 
\app^2 \, \frac{41 q^2}{162} + {\ldots}  \,.
\end{equation}
The representation for the imaginary part just above the 
cut is even more compact,
\begin{multline}
\label{ImPi2R}
{\rm Im} \, \Pi^{(2)}_{\rm R}(q^2 + \ii\epsilon) = 
\frac{\alpha^2}{\pi} \Theta(q^2 - 4) \\
\times \biggl\{ 
\frac{ v^4-2 v^2 - 3}{6} \biggl[ 
\Phi_1\left(\frac{v-1}{v+1}\right) \ln\left(\frac{1-v}{1+v}\right)
\\
-2 \Phi_2\left(\frac{v-1}{v+1}\right) \biggr] 
+ \frac{3v - v^3}{6} \biggl[ 2 \Phi_1\left(\frac{v-1}{v+1}\right)
+3 \ln\left(\frac{1-v}{1+v}\right)\biggr] \\
+ \frac{1}{48} \left(7 v^4-22 v^2-33\right) \ln\left(\frac{1-v}{1+v}\right)
+ \frac{5 v -3 v^3}{8} \biggr\} \,.
\end{multline}

%
%
\subsection{Comparison with the Literature}
\label{comparison}

An essential ingredient of our calculations 
is the master integrals, which should be 
compared to results communicated 
in Refs.~\cite{Da2000,DaKa2001,DaKa2004}.
In Refs.~\cite{Da2000,DaKa2001}, the authors 
define a master integral $J_{011}$ which is equivalent
to our $M_{12}$ up a factor
\begin{equation}
J_{011}= \ii  \, \pi^{2 - \epsilon} \, \Gamma(1 + \epsilon) M_{12} \,.
\end{equation}
In Eq.(1) of Ref.~\cite{Da2000},
the general term of the $\epsilon$-expansion of $J_{011}$
is given in terms of 
log-sine integrals
which can be written in terms of 
Nielsen polylogarithms;
the analytic continuation is shown in Eqs.~(2.9)---(2.23) of 
Ref.~\cite{DaKa2001}.
Moreover,  an closed analytical expression for $J_{011}$ is given in
Eqs.~(2.10)---(2.14) of Ref.~\cite{DaKa2001}, containing 
hypergeometric ${}_2 F_1$ functions.

Changing the variable from  $t = q^2$ to $v$ and using
the transformations for Nielsen polylogarithm 
$S_{n,p}(-1/z) \to S_{n,p}(-z)$,
we found that the results in our Eqs.~\eqref{m12} and~\eqref{n12} 
agree perfectly with the corresponding
terms of Eq.~(1) of Ref.~\cite{Da2000} and the expansion 
given in Eqs.~(2.10)---(2.14) of Ref.~\cite{DaKa2001}.

In Eqs.~(4.9), (4.10), and (E.4) of Ref.~\cite{DaKa2004},
analytical results for master integrals
equivalent to our $M_{235}$ and $M_{235k}$ were presented.
These master integrals are related to ours 
by the following relations: one finds 
for the ultraviolet-convergent integral (4.9) of Ref.~\cite{DaKa2004},
\begin{multline}
4 q^2 (q^2 - 4)^2 J_{011}(1,2,2) = 
-24 (D-3) (D-2) \left(q^2+2\right)
\\
 \times M_{235k}
-(D-2)  \left((D-2) q^4-2 D q^2+16 D-40\right)M_{24}
\\
-4 (D-3)  \left(2 (D-2) q^4-(D-6) q^2+8 D-20\right)M_{235}
\ ,
\end{multline}
while, for the ultraviolet-convergent
linear combination of integrals given in Eq.~(4.10) of Ref.~\cite{DaKa2004},
one has 
\begin{multline}
4 (D - 4) q^2 (q^2 - 4)^3 \left( J_{011}(1,2,2)+2J_{011}(2,1,2)\right)= \\
-(D-2)  \biggl[
(D-4) (D-2) q^6+2 (D-6) (D-4) q^4 \\
+16 (D (5 D-31)+47) q^2 -32 (2 D-7) (2 D-5) \biggr]M_{24} \\
   -4 (D-3) \biggl[ 4 (D-4) (D-2) q^6+3 D (3 D-10) q^4
   \\ +24 (D-4) (D-3) q^2-16 (2 D-7) (2 D-5)
   \biggr] M_{235} \\
-48 (D-3) (D-2) \left(q^2-1\right)  \\
\times \left(D \left(q^2+8\right) -4 \left(q^2+7\right)\right)
M_{235k} \,,
\end{multline}
and for the ultraviolet divergent integral (E.4) of Ref.~\cite{DaKa2004},
\begin{equation}
J_{011}(1,1,1)=M_{235} \,.
\end{equation}
The $\epsilon-$expansion obtained from those of $M_{235}$ and $M_{235k}$
agree perfectly with those presented in Ref.~\cite{DaKa2004}.

Our results for the complete two-loop 
function given in Eq.~\eqref{Pi2R} and for its
imaginary part, as given in Eq.~\eqref{ImPi2R}, agree with the 
calculation of Ref.~\cite{KaSa1955}.
One notes that Eq.~(49) of Ref.~\cite{KaSa1955}
contains the imaginary part,
while Eq.~(57) of Ref.~\cite{KaSa1955} 
also contains the real part 
of the two-loop vacuum
polarization. One should note, though,
that some integrals are left unevaluated
in
Ref.~\cite{KaSa1955}. Furthermore, it needs to 
be pointed out that the expressions in 
Ref.~\cite{KaSa1955}
also contain the reducible diagram 
(see diagram 3 in Fig.~\ref{fig1}) with two 
iterated one-loop vacuum
polarizations on the same line.
Our result for the imaginary part
also agrees with the 
result given in Eq.~(5-4.200) of Ref.~\cite{Sc1970vol3}.
Our results are also in agreement with the 
paper~\cite{BrFlTa1993},
specifically, with the results 
given in Eqs.~(71), (78), (79), (80), and~(81) 
of Ref.~\cite{BrFlTa1993}.
Furthermore, we can refer to related 
investigations on the vector part of the 
quantum chromodynamic vacuum-polarization
tensor~\cite{Kn1990,DjGa1994,ChKuSt1996}.

In Ref.~\cite{BaRe1973}, the two-loop vacuum polarization was
also calculated. The result given in 
Eq.~(49) of Ref.~\cite{BaRe1973} is expressed in 
terms of the variable $\theta$, which in our notation
reads as $\theta = - \frac{1 - v}{1 + v}$.
The last-but-one line of 
Eq.~(49) of Ref.~\cite{BaRe1973} contains two terms,
which, in the notation of Ref.~\cite{BaRe1973},
read as $2 \ln(\theta) \Li_2(\theta)$ and
$2 \ln(\theta) \Li_2(-\theta)$.
The term $2 \ln(\theta) \Li_2(-\theta)$ 
contains a misprint; 
it should have read as $4 \ln(\theta) \Li_2(-\theta)$.
Formula (49) of Ref.~\cite{BaRe1973}
was referred to in Ref.~\cite{Bo1975zpa},
and it was stated in Ref.~\cite{Bo1975zpa}
that Eq.~(49) of Ref.~\cite{BaRe1973}
has a misprint. Indeed, Eq.~(49) of Ref.~\cite{BaRe1973} 
is rewritten as Eq.~(6) of Ref.~\cite{Bo1975zpa}, with a change of
notation: The variable $\delta$ used in Ref.~\cite{Bo1975zpa}
is related to the variable $\theta$ 
used in Eq.~(49) of Ref.~\cite{BaRe1973}  
by the relation $\delta = (1+\theta)/(1-\theta) = v$.
In Eq.~(6) of Ref.~\cite{Bo1975zpa}, 
the wrong term $2 \ln(\theta) \Li_2(-\theta)$ is corrected to 
the right $4 \ln(\theta) \Li_2(-\theta)$, but unfortunately new
misprints are inserted.
The terms (in the notation adapted in Ref.~\cite{BaRe1973})
\begin{subequations}
\begin{align}
T_1 =& \;  -(\delta (5 - 3 \delta^2)/8) \ln \theta \,,
\\[0.1133ex]
T_2 =& \; + \tfrac14 \ln^2\theta 
\left[ \frac{33 + 22 \delta^2 - 7 \delta^4}{24} - \delta (3 - \delta^2) \right] \,,
\end{align}
\end{subequations}
have the wrong signs and should have read 
\begin{subequations}
\begin{align}
T_1 \to & \;
{\widetilde T}_1 =  +(\delta (5 - 3 \delta^2)/8) \ln \theta \,,
\\[0.1133ex]
T_2 \to & \;
{\widetilde T}_2 = -\tfrac14  \ln^2\theta 
\left[ \frac{33 + 22 \delta^2 - 
7 \delta^4}{24} - \delta (3 - \delta^2) \right] \,.
\end{align}
\end{subequations}
In Ref.~\cite{BoRi1982}, Eq.~(6) of Ref.~\cite{Bo1975zpa} is
rewritten as Eq.~(140) of Ref.~\cite{BoRi1982}, 
but the two terms involving $\ln(\theta)$'s keep the wrong sign.

In Ref.~\cite{Pa1993pra}, 
the work from Ref.~\cite{KaSa1955} is being
cited for the imaginary part of the proper two-loop vacuum 
polarization, and it is expressed
in terms of the variable $d$, which equals our $v$.
However, Eq.~(16) of 
Ref.~\cite{Pa1993pra} contains a misprint.
Namely, the term $+3/2 \ln((1+d)/(1-d))$ (in Ref.~\cite{Pa1993pra})
has a logarithm missing and should be read 
$+3/2 \ln((1+d)/(1-d))\ln((1+d)/2)$.
We should clarify that,
despite the typographical error in Eq.~(16) of
Ref.~\cite{Pa1993pra},
final results for the energy shifts 
due to the two-loop vacuum polarization
obtained in Ref.~\cite{Pa1993pra} are in 
agreement with those reported here.

%
%
\section{Contributions to Energy Shifts}
\label{sec4}

%
%
\subsection{Muonic Hydrogen}
\label{sec4A}

Armed with the compact expressions given
in Eqs.~\eqref{Pi2R} and~\eqref{ImPi2R} 
for the imaginary part of the two-loop 
vacuum-polarization function, 
we can evaluate energy shifts in 
hydrogenlike ions. 
In Chap.~10 of Ref.~\cite{JeAd2022book},
the calculation of the vacuum-polarization
energy shift is outlined in detail;
specifically, we refer to Eqs.~(10.244) 
and Eq.~(10.245) of Ref.~\cite{JeAd2022book}.
(We note the different sign conventions
for the sign of the imaginary part of the 
renormalized scalar function $ \Pi_\rmR $,
as compared to Ref.~\cite{JeAd2022book}.)
From the one-loop and two-loop scalar 
functions, one infers the one-loop and 
two-loop irreducible vacuum-polarization potentials
$V^{(1)}_\rmR(r)$ and $V^{(2)}_\rmR(r)$, respectively,
\begin{align}
\label{V1R}
V^{(1)}_\rmR(r) =& \; - \frac{Z \alpha}{\pi} 
\int\limits_{4}^\infty \frac{\dd (q^2)}{q^2}
\; \frac{\ee^{- q r}}{r}
\; \mathrm{Im} \! \left[ \Pi^{(1)}_\rmR (q^2 + \ii \epsilon) \right] \,.
\\
\label{V2R}
V^{(2)}_\rmR(r) =& \; - \frac{Z \alpha}{\pi}
\int\limits_{4}^\infty \frac{\dd (q^2)}{q^2}
\; \frac{\ee^{- q r}}{r}
\; \mathrm{Im} \! \left[ \Pi^{(2)}_\rmR (q^2 + \ii \epsilon) \right] \,.
\end{align}
Of particular phenomenological importance
for the proton radius puzzle~\cite{Je2022proton}
is the contribution of the irreducible two-loop
vacuum-polarization effect on the 
$2P$--$2S$ energy interval in muonic hydrogen.
We use the following results,
\begin{subequations}
\begin{align}
\langle 2S | \frac{\exp(-q \, r)}{r} | 2S \rangle =& \;
\frac{1 + 2 (\beta q)^2}{4 \beta \, ( 1 + \beta q )^4} \,,
\\[0.1133ex]
\langle 2P | \frac{\exp(-q \, r)}{r} | 2P \rangle =& \;
\frac{1}{4 \beta \, ( 1 + \beta q )^4} \,.
\end{align}
\end{subequations}
Here, the $\beta$ parameter is 
\begin{equation}
\beta = \frac{1}{Z\alpha \mu} \,,
\end{equation}
where $Z$ is the nuclear charge number, $\alpha$
is the fine-structure constant, and $\mu$ is the 
reduced mass of the system. For muonic hydrogen,
one has $Z=1$. In units where the electron 
mass is $m = 1$ and the muon mass is $m_\mu$, 
one has $\mu = m_p/(1 + m_p)$ for ordinary hydrogen
and $\mu = m_\mu m_p/(m_\mu + m_p)$ for 
muonic hydrogen.

The contribution of the 
irreducible two-loop vacuum polarization diagrams
to the $2P$--$2S$ splitting in muonic hydrogen is obtained as
\begin{multline}
E^{(2)}(2P-2S) = \app^2 (Z\alpha)^2 \mu f^{(2)}(\beta)\,,
\\
f^{(2)}(\beta) = \int_4^{\oo} \frac{\dd(q^2)}{q^2} 
\frac{{\rm Im} \, \Pi^{(2)}_\rmR (q^2 + \ii \epsilon)}{\alpha^2/\pi} 
\frac{(\beta q)^2}{2 (1+\beta q)^4} \,.
\end{multline}
Quite surprisingly, the two-loop energy shift can be 
expressed analytically. We define
\begin{equation}
\label{wdef}
\beta=\frac{\sqrt{1-w^2}}{2}\ , \qquad w=\sqrt{1-4 \beta^2}\ ,
\end{equation}
and find the result
\begin{align}
\label{f2analytic}
f^{(2)}&(w)=\frac{1}{288 w^6} \biggl\{
-480 \left(w^2-1\right)^2 w^6 I_1\left(\sqrt{\frac{1-w}{1+w}}\right)
\nonumber\\&   
+128 \left(15 w^6-25 w^4+8 w^2+3\right) w^3 I_{2}\left(\sqrt{\frac{1-w}{1+w}}\right)
\nonumber\\&   
+404 w^8-312 w^6-176 w^4+96 w^2
\nonumber\\&   
-32  \pi  \sqrt{1-w^2} \left(15 w^4-19 w^2+3\right) w^4
 +4 \left(w^2-1\right)\times
\nonumber\\&   
     \left(41 w^6+58 w^4-10 w^2-60\right) w 
     \ln\left(\frac{1-w}{1+w}\right)
\nonumber\\&   
-16  \pi ^2 \left(1-w^2\right)^{3/2}
 \left(15 w^6-5 w^4-2 w^2+6\right)
\nonumber\\&   
 -\left(199 w^{10}-302 w^8+51 w^6-4 w^4+176 w^2-96\right) \times
\nonumber\\&   
     \left[\ln^2\left(\frac{1-w}{1+w}\right) +\pi ^2\right]
\biggr\} \,,
\end{align}
where the expressions $I_j(a)$ with $j=1,2$ read as follows,
\begin{align}
\label{I1}
I_1&(a)=
\frac{1}{4}\biggl[
6 \Phi_3(-a^2) - 4 \Phi_2(-a^2) \ln(a^2)
\nonumber\\&   
+ \Phi_1(-a^2) \ln^2(a^2)
 - 4 \ii \pi \left(\Li_2(\ii a) - \Li_2( -\ii a)\right)
\nonumber\\&   
- \pi^2 \ln(1 + a^2)
 - 4 \pi^2 \ln(1 + a)
+ 3 \zeta(3)
\biggr] \,,
\\
\label{I2}
I_{2}&(a)=
\frac{1}{16}\biggl[
-4\Phi_2(-a^2)+4\Phi_1(-a^2) \ln(a^2)
\nonumber\\&   
+3\left(\ln ^2(a^2) +\pi ^2 \right)
-8 \pi  \arctan (a)
\biggr] \,.
\end{align}
Some of the terms could also be expressed in terms of the 
Legendre $\chi$ function, $\chi_\nu(z) = \tfrac12 \, 
[ \Li_\nu(z) - \Li_\nu(-z) ]$.

For muonic hydrogen ($Z=1$), one obtains for the 
nonrelativistic energy shift due to the irreducible two-loop 
diagrams, using CODATA values
for the relevant physical constants~\cite{MoTaNe2008,TiMoNeTa2021}
(see also Ref.~\cite{LaJe2024suppl}),
\begin{equation}
\label{muH}
\left. E^{(2)}(2P-2S) \right|_{\mbox{$\mu$H}} = 1.25298 \, {\rm meV} \,.
\end{equation}
This is consistent with the literature 
(e.g., Ref.~\cite{Je2011aop1}).
The result~\eqref{muH} confirms the numerical
value of a contribution to the proton radius puzzle 
in muonic hydrogen whose numerical magnitude 
could have potentially contributed 
to an explanation the discrepancy. The confirmation
is obtained based on modern 
quantum-field theoretical methods.

The reducible diagram, according to
Eqs.~\eqref{Erdiff} and~\eqref{frdiff}, leads to
an energy shift of
\begin{equation}
\label{muHr}
\left. E^{\rmr}(2P-2S) \right|_{\mbox{$\mu$H}} = 0.25495 \, {\rm meV} \,.
\end{equation}
For selected individual low-lying atomic reference states,
an analytic integration of the one-loop and 
two-loop energy shifts, for 
arbitrary $\beta$, still in the nonrelativistic 
approximation, is 
presented in Appendix~\ref{appd}.

\begin{table}
\caption{Coefficients from the reducible diagram
are indicated for $S$ and $P$ states.}
\label{table1}
\renewcommand{\arraystretch}{2.6}
\small
\begin{center}
\begin{tabular}{lcccccc}
\hline
\hline
   & 1S & 2S & $2P_{1/2}$ & $2P_{3/2}$ \\
\hline
$B^{\rmr}_{50}$
& $-\dfrac{23 \pi }{567}$
& $-\dfrac{23 \pi }{567}$
&$0$
&$0$
\\
$B^{\rmr}_{60}$
&$\dfrac{289}{4050}$
&$\dfrac{181}{4050}$
&$\dfrac{2}{225}$
&$\dfrac{2}{225}$
\\
\hline
\hline
\end{tabular}
\end{center}
\end{table}

\begin{table*}
\caption{Coefficients from the irreducible diagrams
are indicated for $S$ and $P$ states.}
\label{table2}
\small
\renewcommand{\arraystretch}{2.7}
\begin{center}
\begin{tabular}{l@{\hspace*{0.9cm}}c@{\hspace*{0.9cm}}c@{\hspace*{0.9cm}}%
c@{\hspace*{0.9cm}}c}
\hline
\hline
   & 1S & 2S & $2P_{1/2}$ & $2P_{3/2}$ \\
\hline
$B_{40}^{(2)}$
&$-\dfrac{82}{81}$ &$-\dfrac{82}{81}$ & 0 & 0 \\
$B_{50}^{(2)}$
& 
$\dfrac{15647 \pi }{13230} -\dfrac{25 \pi^2}{63}
+\dfrac{52}{63} \pi  \ln (2)$
& 
$\dfrac{15647 \pi }{13230} -\dfrac{25 \pi^2}{63}
+\dfrac{52}{63} \pi  \ln (2)$
& $0$ & $0$ \\
$B_{60}^{(2)}$
& $-\dfrac{82}{81} \ln\left(\dfrac{1}{2}\right)
  +\dfrac{77\zeta (3)}{72}-\dfrac{358}{75}$
& $\dfrac{77 \zeta (3)}{72} -\dfrac{12727}{2700}$
& $-\dfrac{737}{2700}$ & $-\dfrac{449}{5400}$ \\
$B_{61}^{(2)}$
& $-\dfrac{82}{162}$ & $-\dfrac{82}{162}$ & 0 & 0 \\
\hline
\hline
\end{tabular}
\end{center}
\end{table*}

%
%
\subsection{Relativistic Ordinary Hydrogen}
\label{sec4B}

Given our analytic approach,
we may also also study the 
energy shifts to low-lying $S$ and $P$
states for ordinary (electronic) hydrogen,
with a relativistic reference state.
We focus on higher-order terms 
in the semi-analytic expansion 
in powers and logarithms of $Z\alpha$,
for relativistic Dirac-Coulomb reference states
(see Chap.~8 of Ref.~\cite{JeAd2022book}).

The potential $V^{\rmr}_\rmR$ due to the
reducible diagram is
\begin{equation}
\label{Vreducible}
V^{\rmr}_\rmR(r) =
\frac{Z \alpha}{\pi}
\int\limits_{4}^\infty \frac{\dd (q^2)}{q^2}
\; \frac{\ee^{- q r}}{r}
\;
\mathrm{Im}
\left\{
\! \left[ \Pi^{(1)}_\rmR (q^2 + \ii \epsilon) \right] ^2
\right\}\,.
\end{equation}
We study the energy shifts
\begin{equation}
\Delta E^{\rmr}(nS) = 
\langle \psi | V^{\rmr}_\rmR | \psi \rangle \,,
\end{equation}
where $| \psi \rangle$ is the 
relativistic reference-state wave function,
and $\langle \psi $ stands for the 
Hermitian adjoint (not for the Dirac adjoint 
$\overline \psi = \psi^\dagger \, \gamma^0$).
The relevant semi-analytic expansion is 
as follows,
\begin{multline}
\Delta E^{\rmr} =
\app^2 \frac{(Z\alpha)^4}{n^3} 
\biggl\{
(Z\alpha) B_{50}^{\rmr} + 
(Z\alpha)^2 B_{60}^{\rmr} 
\\
+ (Z\alpha)^3 \left[
B_{71}^{\rmr} \, \ln[ (Z\alpha)^{-2} ] + 
B_{70}^{\rmr} \right]
\biggr\} \,.
\end{multline}
For the $nS$ levels,
the dependence on $n$ is given,
for the reducible diagram,
by the following formulas (see also Table~\ref{table1}),
\begin{subequations}
\begin{align}
\label{brnsa}
B_{50}^{\rmr}(nS) = & \; -\frac{23 \pi }{567} \,,
\\[1.1133ex]
\label{brnsb}
B_{60}^{\rmr}(nS) =& \; \frac{29}{810} 
+\frac{8}{225 n^2} \,,
\\
B_{71}^{\rmr}(nS) =& \; \tfrac{1}{2} B_{50}^{\rmr}(nS) \,,
\\[1.1133ex]
B_{70}^{\rmr}(nS) = & \;
-\frac{23 \pi}{567} W_n 
- \frac{570029 \pi}{7858620}
-\frac{46}{567} \pi  \ln(2)
\nonumber
\\
&
+\frac{1294 \pi }{18711 n^2} \,,
\\[1.1133ex]
\label{Wdef}
W_n =& \;  \ln\left( \frac{n}{2} \right) 
- \Psi(n) - \gamma_E + \frac{1}{n}  \,.
\end{align}
\end{subequations}
Here, $\Psi(x) = \Gamma'(x)/\Gamma(x)$ is the 
logarithmic derivative of the Gamma function.
For the $nP_{1/2}$ levels, one has 
$B^\rmr_{50}(nP) = B^\rmr_{71}(nP) = 0$
and the following results for the other coefficients
for the reducible diagram,
\begin{align}
B_{60}^{\rmr}(nP_{1/2})=&
\frac{8}{675} \frac{ n^2-1 }{ n^2 } \,,
\\
B_{70}^{\rmr}(nP_{1/2})=&
-\frac{983 \pi}{93555} \frac{ n^2-1 }{n^2} \,.
\end{align}
For the $nP_{3/2}$ levels, one has the results
\begin{subequations}
\begin{align}
B_{60}^{\rmr}(nP_{3/2})=& \;
B_{60}^{\rmr}(nP_{1/2}) \,,
\\
B_{70}^{\rmr}(nP_{3/2})=& \; 
-
\frac{127 \pi}{17820} \frac{ n^2-1}{ n^2 } \,.
\end{align}
\end{subequations}
The two-loop energy shift from the irreducible 
diagrams, evaluated on the relativistic wave function,
is
\begin{equation}
\Delta E^{(2)} =
\langle \psi | V^{(2)}_\rmR | \psi \rangle \,,
\end{equation}
where $V^{(2)}_\rmR$ is given in Eq.~\eqref{V2R}.
It gives rise to the following
semi-analytic expansion,
\begin{multline}
\Delta E^{(2)} =
\app^2 \frac{(Z\alpha)^4}{n^3}
\biggl\{
B_{40}^{(2)} +
(Z\alpha) B_{50}^{(2)} 
\\
+ (Z\alpha)^2 \left[
B_{61}^{(2)} \, \ln[ (Z\alpha)^{-2} ] +
B_{60}^{(2)} \right]
\biggr\}
\\
+ (Z\alpha)^3 \left[
B_{71}^{(2)} \, \ln[ (Z\alpha)^{-2} ] +
B_{70}^{(2)} \right]
\biggr\} \,.
\end{multline}
For the irreducible diagrams,
the dependence on $n$ is the following
(see also Table~\ref{table2}):
\begin{subequations}
\begin{align}
B^{(2)}_{40}(nS)=& \; -\frac{82}{81} \,,
\\
B^{(2)}_{50}(nS)=& \; 
\frac{15647 \pi }{13230}
-\frac{25 \pi^2}{63}
+\frac{52}{63} \pi  \ln (2) \,,
\\
B^{(2)}_{60}(nS)=& \;
B^{(2)}_{40}(nS) \, W_n\
  +\frac{77 \zeta (3)}{72}
  -\frac{2311}{405}
  +\frac{1313}{675 n^2} \,,
\\
B^{(2)}_{61}(nS)=& \; \tfrac{1}{2} B^{(2)}_{40}(nS) \,,
\qquad
B^{(2)}_{71}(nS)= \tfrac{1}{2} B^{(2)}_{50}(nS) \,.
\end{align}
The Catalan constant $G = 0.915\,965\dots$ 
enters the $B^{(2)}_{70}$ coefficient,
\begin{align}
B^{(2)}_{70}(nS)=& \;
B^{(2)}_{50} \, W_n + B^{(a)}_{70} + \frac{B^{(b)}_{70}}{n^2} \,,
\nonumber\\
B^{(a)}_{70}=&
 \frac{104}{63} \pi \ln ^2(2)
- \frac{25}{63} \pi ^2 \ln (2)
- \frac{200 \, \pi}{63} \, G 
+ \frac{13 \pi^3}{189}
\nonumber\\
&
+ \frac{79318 \pi  \ln(2)}{19845}
- \frac{82361 \pi^2}{79380}
+ \frac{89515961 \pi }{20003760} \,,
\\
B^{(b)}_{70}=&-\frac{3304831 \pi}{1428840}
+\frac{1891 \pi ^2}{2268}
-\frac{1076}{567} \pi\ln (2) \,.
\end{align}
\end{subequations}
For the $nP_{1/2}$ levels,
the coefficients are as follows,
\begin{subequations}
\begin{align}
B^{(2)}_{60}(nP_{1/2})=&-\frac{737}{2025} 
\frac{n^2-1 }{ n^2 } \,,
\\
B^{(2)}_{70}(nP_{1/2})=& \pi
\biggl[
 \frac{1764797}{7144200}
-\frac{809 \pi }{11340}
+\frac{328 \ln (2)}{2835}
\biggr] \frac{n^2-1}{n^2} \,.
\end{align}
\end{subequations}
For the $nP_{3/2}$ levels,
the coefficients are as follows,
\begin{subequations}
\begin{align}
B^{(2)}_{60}(nP_{3/2})=& \; 
-\frac{449}{4050} \, \frac{n^2-1}{n^2}\ ,
\\
B^{(2)}_{70}(nP_{3/2})=& \pi 
\left[\frac{75763}{510300}-\frac{31 \pi }{810}+\frac{19 \ln
(2)}{405}\right]
\frac{ n^2-1 }{ n^2 } \,.
\end{align}
\end{subequations}
The fine-structure difference 
\begin{equation}
B^{(2)}_{60}(nP_{3/2}) - 
B^{(2)}_{60}(nP_{1/2})= \frac{41}{162}
\frac{n^2-1 }{ n^2 }
\end{equation}
is consistent with the result communicated in 
Eq.~(7.7) of Ref.~\cite{JeCzPa2005}.

%
%
\section{Conclusions}
\label{sec5}

We have completed a complete reevaluation
of the two-loop vacuum-polarization
tensor in dimensional regularization,
on the basis on integration-by-parts
identities.
The two-loop vacuum-polarization
tensor constitutes a numerically significant
contribution to the Lamb shift of muonic hydrogen which
influences the determination of the proton radius 
from muonic hydrogen spectroscopy~\cite{Je2011aop1}.

In Sec.~\ref{sec2}, we have discussed the evaluation 
of the one-loop vacuum polarization insertion into
the photon propagator, evaluated to 
order $\epsilon$, and $\epsilon^2$, and
thus, in a form suitable for inclusion 
into higher-order loop calculation where 
knowledge of the terms of higher orders in
$\epsilon$ is indispensable. 
We note that we use somewhat nonstandard 
conventions for the $\overline{\mathrm{MS}}$-charge,
as given in  Eq.~\eqref{id}.

In Sec.~\ref{sec3}, the irreducible two-loop vacuum polarization
insertion has been evaluated, by expressing it in terms
of master integrals. 
The renormalization has been carried out,
and final results have been presented for the 
real and imaginary parts, in Eqs.~\eqref{Pi2R} and~\eqref{ImPi2R}.
A comparison to the existing literature is being
performed as well (Sec.~\ref{comparison}).

In Sec.~\ref{sec4} and Appendix~\ref{appd}, we demonstrate that,
for arbitrary reduced mass of a two-body 
bound system, the two-loop vacuum-polarization corrections 
to the energy can be evaluated analytically
(for nonrelativistic reference states)
and expressed in terms of dilogarithmic, and
trilogarithmic, functions.
This applies both to the $2P$--$2S$ difference
(see Sec.~\ref{sec4A}) as well as 
to individual hydrogenic levels 
(see Appendix~\ref{appd}).
Higher-order coefficients for the semi-analytic 
expansion of the two-loop vacuum-polarization
energy shift could be  evaluated with the 
help of the fully relativistic
hydrogen wave function. Results have been presented in 
Sec.~\ref{sec4B}.

\section*{Acknowledgments}

The authors acknowledge insightful
conversations with Professor Ettore Remiddi.
Support from the National Science Foundation
(grant PHY--2110294) is gratefully acknowledged.
S.L. also acknowledges support from Italian Ministry of University and
Research (MUR) via the PRIN 2022 project n.~20225X52RA --- MUS4GM2
funded by the European Union via the Next Generation EU package.

\appendix

%
%
\section{Terms of Order $\maybebm{\epsilon^2}$}
\label{appa}

We first give the terms quadratic in $\epsilon$ for Eq.~\eqref{m12},
for the master integral $M_{12}$,
\begin{widetext}
\begin{align}
\label{n12}
N_{12}(v)&=
- 12 \text{Li}_3\left(\frac{1-v}{v+1}\right) 
- 24 \text{Li}_3\left(\frac{2 v}{v+1}\right) 
+ \ln^3\left(\frac{v-1}{v+1}\right)
- 6 \ln \left(\frac{2 v}{v+1}\right) 
\ln^2\left(\frac{v-1}{v+1}\right)
\nonumber\\ 
& +12 \ln^2\left(\frac{2 v}{v+1}\right)
\left( \ln\left(\frac{v-1}{v+1}\right)
- \ln\left(-\frac{v-1}{v+1}\right) \right)
- \pi ^2 \ln\left(\frac{v-1}{v+1}\right)
+ 6 \pi ^2 \ln\left(\frac{2 v}{v+1}\right)
+12 \, \zeta (3) \,.
\end{align}
We now give the expressions for the terms of order $\epsilon^2$ 
in the one-loop effect, as discussed in Eq.~\eqref{P1R}. For the 
quadratic term $Q^{(1)}_2(v)$ in Eq.~\eqref{P1R},
one obtains the result 
\begin{align}
\label{P1Rquadratic}
Q^{(1)}_2(v) =& \;
\left(v-\frac{4 v^3}{9}\right) 
\biggl[ -2 \text{Li}_2\left(\frac{1-v}{v+1}\right)
+ \frac{1}{2} \ln^2\left(\frac{v-1}{v+1}\right)
- 2 \ln \left(\frac{2v}{v+1}\right) \ln \left(\frac{v-1}{v+1}\right)
-\frac{\pi^2}{6} \biggr]
\nonumber\\ 
& + \frac{1}{36} v \left(v^2-3\right) 
\biggl[ 12 \text{Li}_3\left(\frac{1-v}{v+1}\right)
+ 24 \text{Li}_3\left(\frac{2 v}{v+1}\right)
- \ln^3\left(\frac{v-1}{v+1}\right)
+ 6 \ln \left(\frac{2 v}{v+1}\right) 
   \ln^2\left(\frac{v-1}{v+1}\right)
\nonumber\\ 
& + 12 \left( \ln\left(\frac{1-v}{v+1}\right)
- \ln \left(\frac{v-1}{v+1}\right)\right) 
\ln^2\left(\frac{2 v}{v+1}\right)
+ \pi^2 \ln\left(\frac{v-1}{v+1}\right)
- 6 \pi^2 \ln\left(\frac{2 v}{v+1}\right)
- 12 \zeta (3) \biggr]
\nonumber\\ 
& - \frac{2}{81} \left(\left(39 v^3-81 v\right) 
\ln \left(\frac{v-1}{v+1}\right) + 78 v^2 - 160 \right) \,.
\end{align}
For the term $Q_2^{(1a)}(v)$ 
from the integral with two fermion propagators, one finds
for the term quadratic in $\epsilon$ the expression
\begin{align}
Q^{(1a)}_2(v) =& \; \frac{v^2 - 1}{4v} \,
\biggl\{ -\frac{1}{6} \left(v^2-1\right) 
\biggl[12 \text{Li}_3\left(\frac{1-v}{v+1}\right)
+ 24 \text{Li}_3\left(\frac{2 v}{v+1}\right)
- \ln ^3\left(\frac{v-1}{v+1}\right)
+6 \ln \left(\frac{2v}{v+1}\right) 
\biggl( \ln^2 \left(\frac{v-1}{v+1}\right) -\pi^2 \biggr)
 \nonumber\\ 
& -24 \ln^2\left(\frac{2 v}{v+1}\right) 
\ln \left(\frac{v-1}{v+1}\right)
+ 12 \left(\ln\left(\frac{1-v}{v+1}\right)
+ \ln \left(\frac{v-1}{v+1}\right)\right) 
\ln^2\left(\frac{2 v}{v+1}\right)
+ \pi ^2 \ln \left(\frac{v-1}{v+1}\right)
\nonumber\\ 
& -12 \zeta (3) \biggr] 
+v^2 \biggl[ -4 \text{Li}_2\left(\frac{1-v}{v+1}\right)
+ \ln ^2\left(\frac{v-1}{v+1}\right) 
- 4 \left(\ln \left(\frac{2 v}{v+1}\right)-1\right) 
\ln\left(\frac{v-1}{v+1}\right)
- \frac{\pi ^2}{3} \biggr] +8 v \biggr\} \,.
\end{align}   
%
%
\section{Coefficient Functions}
\label{appb}

The coefficients in Eq.~\eqref{redu21} are as follows,
\begin{align}
\calP^{(2;1)}_{234} =& \;
-\frac{(2-D) \left(D^2 \left(-q^2-4\right)-D \left(q^4-5 q^2-28\right)+2
   \left(q^4-2 q^2-20\right)\right)}{2 (D-4) (D-3) (D-1) q^2
   \left(4-q^2\right)}
  \,,
\\
\calP^{(2;1)}_{24} =& \;
-\frac{(D-2) \left(3 D^2 q^4-4 D \left(3 q^4-2
   q^2+8\right)+4 \left(3 q^4-8 q^2+20\right)\right)}{2 (D-4) (D-3) (D-1) q^4
   \left(4-q^2\right)^2}
  \,,
\\
\calP^{(2;1)}_{1234} =& \;
\frac{-\left(D^3-9 D^2+30 D-32\right)
q^4+4 \left(D^3-11 D^2+38 D-40\right) q^2+32}{8 (D-4) (D-1) q^2 
\left(4-q^2\right)}
\,,
\\
\calP^{(2;1)}_{235k} = & \;
-\frac{ 1 }{(D-4) (D-3) (D-1) q^4 \left(4-q^2\right)^2} \biggl[
3 \left(-4 D^3 q^2 \left(1-q^2\right)-D^2
   \left(-q^6+30 q^4-36 q^2+16\right) \right.
\nonumber\\
& \left. +4 D \left(-q^6+18 q^4-28 q^2+20\right)-4
   \left(-q^6+14 q^4-28 q^2+24\right)\right)
\biggr] \,,
\\
\calP^{(2;1)}_{235} = & \;
-\frac{ 1 }{4 (D-4) (D-3) (D-1) q^4 \left(4-q^2\right)^2}
\biggl[
 - D^4 q^2 \left(4-q^2\right)^2+D^3 q^2 
 \left(23 q^4-100  q^2+176\right)
\nonumber\\
& +4 D^2 \left(q^8-35 q^6+119 q^4-176 q^2-32\right)
  -8 D \left(2 q^8-40 q^6+127 q^4-148 q^2-88\right)
\nonumber\\
&  +8 \left(2 q^8-31 q^6+98 q^4-84 q^2-120\right)
\biggr] \,.
\end{align}
The coefficients in Eq.~\eqref{redugauge} are as follows,
\begin{align}
\calG^{(2;1)}_{235k} =& \;
-\frac{3 (D-2) \left(-D^2 \left(-q^2-2\right) q^2 
  - D \left(4 q^4 + 6 q^2 + 8\right) +
  4 \left(q^4-q^2+6\right)\right)}{(D-4) (D-1) q^4 \left(4-q^2\right)^2} \,,
\\
\calG^{(2;1)}_{24} =& \;
 \frac{(2-D) \left(3 D^2 q^4-4 D \left(3 q^4-2 
 q^2+8\right)+4 \left(3 q^4-8 q^2+20\right)\right)}{4 (D-4) (D-1) q^4 
 \left(4-q^2\right)^2} \,,
\\
\calG^{(2;1)}_{235} =& \;
 \frac{ 1 }{4 (D-4) (D-1) q^4 \left(4-q^2\right)^2}
 \biggl[ - 3 D^3 \left( q^2 + 2 \right) q^4+D^2 
 \left(18 q^6+26 q^4+64\right)
\nonumber\\
&  +4 D \left(-9 q^6-4 q^4+8 q^2-88\right)+8 \left(3 
   q^6-2 q^4-10 q^2+60\right) \biggr] \,.
\end{align}
The coefficients in Eq.~\eqref{redu22} are as follows,
\begin{align}
\calP^{(2;2)}_{234} =& \;
\frac{(D-2) \left(-2 D^2 q^2-D \left(q^4-14 q^2+8\right)+2 \left(q^4-10
   q^2+12\right)\right)}{16 (D-3) q^2 \left(4-q^2\right)}  \,,
\\
\calP^{(2;2)}_{235k} =& \;
\frac{3 (D-2) }{8 (D-4) (D-3) (D-1) q^4 \left(4-q^2\right)^2}
\biggl[ - D^3 q^2 \left(4-q^2\right)^2+D^2 q^2 \left(7 q^4-48
   q^2+128\right)
\nonumber\\
&  +2 D \left(-7 q^6+40 q^4-136 q^2-32\right)
   -8  \left(-q^6+4 q^4-12 q^2-24\right) \biggr] \,,
\\
\calP^{(2;2)}_{235} =& \;
-\frac{ 1 }{8 (D-4) (D-3) (D-1) q^4 \left(q^2-4\right)^2}
\biggl[ D^4 q^2 \left(q^2-4\right)^2 
       \left(q^2+1\right)-3 D^3 q^2 \left(3 q^6-19 q^4
\right.
\nonumber\\
& \left.
   +28 q^2+48\right)+4 D^2 \left(7
    q^8-40 q^6+69 q^4+112 q^2+32\right)-4 D \left(9 q^8-45 q^6+80 q^4
   \right.
\nonumber\\
&\left.
   +128 q^2+176\right)
   +16 \left(q^8-4 q^6+6 q^4+6 q^2+60\right) \biggr] \,,
\\
\calP^{(2;2)}_{24} =& \;
\frac{D-2}{4 (D-4) \left(D^2-4 D+3\right) q^4 \left(q^2-4\right)^2}
\biggl[ D^4 q^2 \left(q^2-4\right)-8 D^3 q^2 \left(q^2-4\right)
\nonumber\\
& +12 D^2 q^2 \left(2 q^2-7\right)+
    D \left(-34 q^4+96 q^2-32\right)+20 q^4-64 q^2+80 \biggr] \,.
\end{align}
%
%
\section{Higher--Order Terms in the Master Integrals}
\label{appc}

We have the following results for the higher-order
terms from the master integrals, with reference
to Eqs.~\eqref{m235} and~\eqref{m235k}.
For $N_{235}(q^2)$, we obtain
\begin{align}
N_{235}(q^2)&=
\frac{1}{6} v \left(v^2-3\right) \biggl[
 18
   \Phi_2\left(\frac{v-1}{v+1}\right)-18 \ln \left(\frac{v-1}{v+1}\right)
   \Phi_1\left(\frac{v-1}{v+1}\right)-6
   \text{Li}_2\left(\frac{v-1}{v+1}\right)
   -6 \ln \left(\frac{2}{v+1}\right) \ln
   \left(\frac{v-1}{v+1}\right)
      \\\nonumber &
   -39 \ln \left(\frac{v-1}{v+1}\right)+\pi
   ^2\biggr]
   +\frac{1}{12} \left(v^4+2 v^2-3\right) \biggl[
   36
   \Phi_3\left(\frac{v-1}{v+1}\right)-18 \ln \left(\frac{v-1}{v+1}\right)
   \Phi_2\left(\frac{v-1}{v+1}\right)
      \\\nonumber &
   -6
   \text{Li}_2\left(\frac{v-1}{v+1}\right) \ln \left(\frac{v-1}{v+1}\right)-3
   \ln ^3\left(\frac{v-1}{v+1}\right)-6 \ln \left(\frac{2}{v+1}\right) \ln
   ^2\left(\frac{v-1}{v+1}\right)+\pi ^2 \ln \left(\frac{v-1}{v+1}\right)+18
   \zeta (3)
   \biggr]
      \\\nonumber &
   +\frac{1}{8} \left(60 v^2+\left(-5 v^4-16 v^3-14 v^2+48
   v+3\right) \ln ^2\left(\frac{v-1}{v+1}\right)+170\right) \,.
\end{align}
For $N_{235k}(q^2)$, we obtain
\begin{align}
  N_{235k}(v)&=
\frac{1}{36} v \left(3 v^4-8 v^2+21\right) \biggl[
18 \Phi_2\left(\frac{v-1}{v+1}\right) 
- 18 \ln \left(\frac{v-1}{v+1}\right) \Phi_1\left(\frac{v-1}{v+1}\right)
- 6 \text{Li}_2\left(\frac{v-1}{v+1}\right)
\\\nonumber 
& -6 \ln \left(\frac{2}{v+1}\right) 
\ln\left(\frac{v-1}{v+1}\right)+\pi^2 \biggr] 
+ \frac{1}{24} \left(v^6-3 v^4-5 v^2+7\right) 
\biggl[ 36 \Phi_3\left(\frac{v-1}{v+1}\right) 
- 18 \ln\left(\frac{v-1}{v+1}\right) 
\Phi_2\left(\frac{v-1}{v+1} \right)
\nonumber \\
& -6 \text{Li}_2\left(\frac{v-1}{v+1}\right) \ln \left(\frac{v-1}{v+1}\right)
-3 \ln^3\left(\frac{v-1}{v+1}\right)-6 \ln \left(\frac{2}{v+1}\right) 
\ln^2\left(\frac{v-1}{v+1}\right) 
+ \pi^2 \ln \left(\frac{v-1}{v+1}\right) + 18 \zeta (3) \biggr]
\nonumber\\
& +\frac{1}{48} \biggl[ -2 \left(86 v^4+233 v^2+601\right)
- 4 v \left(37 v^4-108 v^2+279\right) \ln \left(\frac{v-1}{v+1}\right)
\nonumber \\
& +\biggl( -13 v^6-48 v^5+63 v^4+128 v^3 +105 v^2-336 v-27 \biggr) 
\ln^2\left(\frac{v-1}{v+1}\right) \biggr] \,.
\end{align}   
\end{widetext}
  
%
%
\section{Analytic Integrations}
\label{appd}

%
%
\subsection{One--Loop Diagram}

We recall the imaginary part of the one-loop effect
from Eq.~\eqref{ImPi1R},
\begin{equation}
\mathrm{Im} \! \left[ \Pi^{(1)}_\rmR(q^2 + \ii \epsilon) \right]
= \frac{\alpha}{3} \sqrt{1- \frac{4}{q^2}} \,
\left( 1 + \frac{2}{q^2}\right) \,,
\end{equation}
for $q^2 \geq 4$.
We use Eq.~(10.245) of Ref.~\cite{JeAd2022book}
and write
\begin{equation}
V^{(1)}_R(r) = - \frac{Z \alpha}{\pi} 
\int\limits_{4}^\infty \frac{\dd (q^2)}{q^2}
\; \frac{\ee^{- q r}}{r}
\; \mathrm{Im} \! \left[ \Pi^{(1)}_\rmR (q^2 + \ii \epsilon) \right] \,.
\end{equation}
One uses the well-known formula for the 
generalized Bohr radius in a hydrogenlike 
ion with nuclear charge number $Z$, namely,
$\beta \equiv a_0 = 1/(Z\alpha \mu)$,
where $\mu$ is the reduced mass of the 
two-body system, measured in units of the 
electron mass.
For the $2S$ and $2P$ states of atomic hydrogen,
one finds, respectively,
\begin{subequations}
\begin{align}
\langle 2S | \frac{\exp(-q \, r)}{r} | 2S \rangle =& \;
\frac{1 + 2 (\beta q)^2}{4 \beta \, ( 1 + \beta q )^4} \,,
\\[0.1133ex]
\langle 2P | \frac{\exp(-q \, r)}{r} | 2P \rangle =& \;
\frac{1}{4 \beta \, ( 1 + \beta q )^4} \,.
\end{align}
\end{subequations}
Hence, 
\begin{multline}
\langle 2P | \frac{\exp(-q \, r)}{r} | 2P \rangle -
\langle 2S | \frac{\exp(-q \, r)}{r} | 2S \rangle \\
= - \frac{(\beta q)^2}{2 \, \beta \, ( 1 + \beta q )^4} \,,
\end{multline}
The contribution to the $2P$--$2S$ energy shift is
\begin{multline}
E_1(2P-2S)= \langle 2P | V^{(1)}_R | 2P \rangle -
\langle 2S | V^{(1)}_R | 2S \rangle 
\\
= \frac{(Z \alpha)^2 \mu}{2 \pi}
\int\limits_{4}^\infty \frac{\dd (q^2)}{q^2}
\; \mathrm{Im} \! \left[ \Pi_\rmR^{(1)} (q^2 + \ii \epsilon) \right] \,
\frac{(\beta q)^2}{( 1 + \beta q )^4} 
\\
= \app \, (Z\alpha)^2 \mu f_1(\beta)\,,
\\
f_1(\beta) = \int_4^{\oo} \frac{\dd(q^2)}{q^2}
\frac{{\rm Im} \Pi_\rmR^{(1)} (q^2 + \ii \epsilon)}%
{\alpha} \frac{(\beta q)^2}{2 (1+\beta q)^4} \,.
\end{multline}
We choose the $2P$--$2S$ interval in order to 
demonstrate that energy shifts due to the 
one-loop and two-loop vacuum-polarization
effects can be evaluated analytically.
For muonic hydrogen, one has the result $\mu = m_\mu m_p/(m_\mu + m_p)$,
where $m_\mu$ and $m_p$ are the muon and proton masses,
respectively.

It turns out to be advantageous to express the 
result in terms of the variable $w$ defined in 
Eq.~\eqref{wdef}, with the result
\begin{align}
f_1(w)=&
\frac{\left(w^2-1\right)^2}{72 w^5}
\biggl[
 3
\left(8 w^4+4 w^2+3\right) \ln \left(\frac{1-w}{1+w}\right)
\nonumber \\&
+\frac{24 \pi  w^5}{\sqrt{1-w^2}}
+\frac{2 \left(24 w^6-28 w^4-3 w^2+9\right) w}{\left(w^2-1\right)^2}
\biggr] \,.
\end{align}
We can get the expansions for small and large $\beta$,
\begin{subequations}
\label{expansion1}
\begin{align}
f_1(\beta)=& \; 
\frac{1}{18}-\beta ^2+\frac{8 \pi  \beta ^3}{3}+\beta ^4 \left(10 \ln
\left(\beta^2\right) +
\frac{32}{3}\right)
\nonumber\\[0.1133ex]
& \; +\beta^6 \left(\frac{140 \ln\left(\beta^2\right)}{3}
+\frac{236}{3}\right) + \calO\left(\beta ^8\right) \,,
\\[0.1133ex]
f_1(\beta)=& \; \frac{1}{30 \beta ^2}-\frac{5 \pi }{384 \beta ^3}+\frac{1}{28 \beta^4}
-\frac{35 \pi }{4096 \beta ^5}+\frac{1}{54 \beta ^6}
\nonumber\\[0.1133ex]
& -\frac{63 \pi }{16384 \beta^7} + \calO\left(\frac{1}{\beta^8 }\right)\, .
\end{align}
\end{subequations}
For muonic hydrogen, one has $\beta = 0.737384\dots$, 
while, for ordinary hydrogen, one has $\beta = 137.110\dots$.
Hence, the second of the above equations
is relevant for ordinary hydrogen,
while none of the expansions can be used with 
good accuracy for muonic hydrogen.
It should be possible, though,
to generalize the results reported in 
this Appendix to reference states other than
$2S$ and $2P$ if desired.
Relativistic corrections to the one-loop
vacuum-polarization shift have been 
analyzed in Ref.~\cite{Je2011pra}, with an 
emphasis on muonic hydrogen.

%
%
\subsection{Two--Loop Reducible Diagram}

Next, we consider the contribution of the reducible diagram~3 in 
Fig.~\ref{fig1},
\begin{multline}
\label{Erdiff}
E^{\rmr}(2P-2S) = \app^2 (Z\alpha)^2 \mu \, f^{\rmr}(\beta)\,,
\\
f^{\rmr}(\beta) = - \int_4^{\infty} \frac{\dd(q^2)}{q^2} 
\frac{ {\rm Im}\left[ \left(
\Pi^{(1)}_\rmR(q^2 + \ii \epsilon)
\right)^2\right] }{\alpha^2/\pi} \,
\frac{(\beta q)^2}{2 (1+\beta q)^4} \,.
\end{multline}
We use the conventions outlined Eq.~\eqref{wdef}.
An analytic evaluation leads to the formula
\begin{multline}
\label{frdiff}
f^{\rmr}(w)= - \frac{1}{648 w^5}
\biggl\{ 4 w \biggl(-315 w^8+1005 w^6-836 w^4
\\ 
+84 w^2 +72\biggr) +12 \biggl(-105 w^{10}+370 w^8-381 w^6
\\   
+96 w^4 +10 w^2+12\biggr) \ln  \left(\frac{1-w}{1+w}\right)
\\   
-45 \left(w^2-1\right)^2\left(7 w^2-13\right) w^5
\biggl[ \ln^2\left(\frac{1-w}{1+w}\right) +\pi ^2
\biggr] \biggr\} \,.
\end{multline}
The expansion for small $\beta$ reads as follows,
\begin{multline}
\label{expansion2rsmall}
f^{\rmr}(\beta)= - \frac{\ln \left(\beta ^2\right)}{27}
-\frac{5}{81}
+\beta ^2 \left(\frac{2}{3} \ln \left(\beta^2\right)
+\frac{32}{9}\right)
\\
-\frac{\beta^4}{27} \left(180 \ln ^2\left(\beta ^2\right)
+774 \ln\left(\beta ^2\right)
+180 \pi ^2-73\right) + \calO(\beta^6)\,.
\end{multline}
For large $\beta$, one obtains the result
\begin{align}
\label{expansion2rlarge}
f^{\rmr}(\beta)=& \;
  \frac{23 \pi }{4536 \beta ^3}
-\frac{1}{45 \beta ^4}
+ \frac{127 \pi}{19008 \beta^5}
-\frac{1}{60 \beta ^6}
\nonumber\\
& \;
+\frac{1675 \pi }{439296 \beta^7}
+ \calO\left(\frac{1}{\beta^8 }\right)  \,.
\end{align}

%
%
\subsection{Two--Loop Irreducible Diagram}

We continue with the contribution of the 
irreducible two-loop vacuum polarization diagrams,
\begin{multline}
E_{2}(2P-2S) = \app^2 (Z\alpha)^2 \mu f_{2}(\beta)\,,
\\
f_{2}(\beta) = \int_4^{\oo} \frac{\dd(q^2)}{q^2} 
\frac{{\rm Im} \, \Pi^{(2)}_\rmR (q^2 + \ii \epsilon)}{\alpha^2/\pi} 
\frac{(\beta q)^2}{2 (1+\beta q)^4} \,.
\end{multline}
The result is a bit more complex and 
given in Eq.~\eqref{f2analytic}.
The expansion for small $\beta$ is
\begin{align}
\label{expansion2small}
f_2&(\beta)=
\frac{1}{24} - \beta ^2 \left(3 \ln (\beta) + \frac{11}{2}\right)
+\beta ^3\left(\frac{320 \pi}{27}-\frac{56 \pi ^2}{9}\right) 
\nonumber\\
& +\beta ^4 \left(60 \ln^2(\beta )+69 \ln (\beta )
 -20 \zeta (3)+15 \pi^2+\frac{43}{12}\right)
\nonumber\\
& + \calO\left(\beta^5\right) \,.
\end{align}
For large $\beta$, one finds the result
\begin{align}
\label{expansion2large}
f_2&(\beta)= \frac{41}{324 \beta ^2} 
+ \frac{1}{\beta^3} \left(-\frac{15647 \pi}{105840} 
+ \frac{25 \pi^2}{504}
- \frac{13}{126} \pi \ln(2) \right)
\nonumber\\
& + \frac{449}{2160 \beta ^4}
+ \frac{1}{\beta^5} \left( -\frac{75763 \pi }{544320}
+ \frac{31 \pi^2}{864} 
- \frac{19}{432} \pi  \ln (2) \right)
\nonumber\\
& + \frac{62479}{453600 \beta ^6}
+ \frac{1}{\beta ^7} \biggl[ -\frac{71317517 \pi }{936714240}
+ \frac{37 \pi ^2}{2112}
\nonumber\\
& - \frac{125}{8448} \pi \ln (2) \biggr]
+ \calO\left( \frac{1}{\beta^8} \right) \,.
\end{align}

%
%
\subsection{Individual Energy Shifts}

In Sec.~\ref{sec4A}, we had concentrated
on the $2P$--$2S$ energy difference,
for muonic hydrogen.
It is instructive to 
separately give the nonrelativistic contributions 
to selected energy levels 
with principal quantum number $n$ 
and orbital angular momentum quantum number $\ell$
(in the nonrelativistic approximation)
from the one-loop diagram, the 
two-loop irreducible and the 
two-loop reducible diagrams,
\begin{subequations}
\begin{align}
E^{(1)}(n\ell) = & \; \app \, (Z\alpha)^2 \mu f^{(1)}_{n\ell}(\beta)\,,
\\[1.1133ex]
E^{\rmr}(n\ell) = & \; \app^2 \, (Z\alpha)^2 \mu f^{\rmr}_{n\ell}(\beta)\,,
\\[1.1133ex]
E^{(2)}(n \ell) =& \; \app^2 \, (Z\alpha)^2 \mu f^{(2)}_{n \ell}(\beta)\,.
\end{align}
\end{subequations}
We consider the cases 
$| n\ell \rangle = | 1S \rangle, |2S\rangle, |2P\rangle$.
The variable $w$ has been defined in Eq.~\eqref{wdef}.
We define the variable $u$ as follows,
\begin{equation}
u=\sqrt{1-\beta^2}\,, \quad \beta=\sqrt{1-u^2} \,.
\end{equation}
For the ground state,
we obtain
\begin{multline}
f^{(1)}_{1S}(\beta)=\frac{1}{18 u }
\biggl[
3 \pi  \sqrt{1-u^2} \left(4 u^2-7\right) u
-24 u^3+46 u  
\\
-3 \left(4 u^4-9
u^2+3\right) \ln \left(\frac{1-u}{1+u}\right)
\biggr] \,.
\end{multline}
For the $2S$ state, one has the result
\begin{multline}
f^{(1)}_{2S}(\beta) =\frac{1}{144 w^5 }
\biggl[
-336 w^7+464 w^5-18 w^3 -54 w
\\
+12 \pi w^5 \left(14 w^2-17\right)  \sqrt{1-w^2}
\\
-3 \left(56 w^8-96 w^6+27 w^4+9\right) \ln
   \left(\frac{1-w}{1+w}\right)
\biggr] \,.
\end{multline}
For the $2P$ level, we obtain
\begin{multline}
f^{(1)}_{2P}(\beta) =\frac{1}{144 w^5 }
\biggl[
-240 w^7+352 w^5-30 w^3 -18 w
\\
+12 \pi w^5  \left(10 w^2-13\right)  \sqrt{1-w^2} 
\\
-3 \left(40 w^8-72 w^6+21 w^4+4 w^2+3\right) \ln
   \left(\frac{1-w}{1+w}\right)
\biggr] \,.
\end{multline}
For the reducible diagram, the results are as follows.
Let us first indicate the result for the ground state,
\begin{multline}
f^{\rmr}_{1S}(\beta) = - 
\frac{1}{1620 u } \biggl[ -288 \pi  u \sqrt{1-u^2}
\\
+20 u \left(63 u^4-282 u^2+304\right)
\\
+45 u \left(7 u^6-36 u^4+51  u^2-18\right)
\left[ \ln^2\left(\frac{1-u}{1+u}\right)+\pi^2 \right]
\\
+60 \left(21 u^6-101 u^4+126 u^2-24\right) \ln
   \left(\frac{1-u}{1+u}\right)
\biggr] \,.
\end{multline}
For the $2S$ state, one has the result
\begin{multline}
\label{fr2S}
f^{\rmr}_{2S}(\beta)=
- 
\frac{1}{6480 w^5 }
\biggl[ -576 \pi w^5 \sqrt{1-w^2}
\\
+20 \left( 1386 w^8-4719 w^6+4132 w^4-396 w^2-216 \right) w
\\
+45  w^5  \left(154 w^6-627 w^4+792 w^2-315\right) 
\left[ \ln^2\left(\frac{1-w}{1+w}\right)+\pi^2\right]
\\
+60 \left(462 w^{10}-1727  w^8+1851 w^6-468 w^4-54 w^2-36\right)
\times
\\
\ln \left(\frac{1-w}{1+w}\right) \biggr] \,,
\end{multline}
while the $2P$ result is
\begin{multline}
\label{fr2P}
f^{\rmr}_{2P}(\beta)=
- \frac{1}{6480 w^5}
\biggl[ -576 \pi w^5 \sqrt{1-w^2} 
\\
+60 \left(252 w^8-903 w^6+820 w^4-76 w^2-24\right) w
\\
+45   w^5   \left(84 w^6-357 w^4+462 w^2-185\right)
\times
\\
 \left[\ln ^2\left(\frac{1-w}{1+w}\right) +\pi^2 \right]
+60 \biggl(
252 w^{10}-987 w^8
\\
+1089 w^6-276 w^4-34 w^2-12
\biggr) 
    \ln \left(\frac{1-w}{1+w}\right)
\biggr] \,.
\end{multline}
The difference of the 
results given in Eqs.~\eqref{fr2P} and~\eqref{fr2S} 
confirms Eq.~\eqref{frdiff}.

The results for the irreducible two-loop diagram
are more complicated and read as follows
[we refer to the definitions of $I_1$ and $I_2$
in Eqs.~\eqref{I1} and~\eqref{I2}].
For the ground state, one obtains
\begin{multline}
f^{(2)}_{1S}(\beta)=\frac{1}{96 u^2}
\biggl[
128 \left(11-5 u^2\right) u^3
      I_2\left(\sqrt{\frac{1-u}{1+u}}\right)
\\      
+\left(43 u^6-94 u^4-45 u^2-48\right) 
\left(\ln^2\left(\frac{1-u}{1+u}\right)+\pi ^2\right)
\\
   +4 \left(43-37 u^2\right) u^3 \ln   \left(\frac{1-u}{1+u}\right)
\\
      +4 u^2 \left(8  \left(5 u^4-10 u^2+1\right)
      I_1\left(\sqrt{\frac{1-u}{1+u}}\right)
      -57 u^2+74\right)
\\      
   +\sqrt{1-u^2} \biggl[
   \pi  \left(\frac{32}{9} \left(45 u^2-113\right) u^2
   +\frac{1024}{3} u^2 \ln(2)\right)
\\   
   +\pi ^2 \left(-80 u^4+\frac{320 u^2}{3}+48\right)
   \biggr]
\biggr] \,.
\end{multline}
For the metastable $2S$ level, 
the analytic integration
leads to the result
\begin{multline}
f^{(2)}_{2S}(\beta)=\frac{1}{1152 w^6}
\biggl[
3 \biggl(
 \biggl(
  653 w^{10}-1058 w^8+93 w^6
  \\
 +56 w^4+256 w^2-192
 \biggr)
 \left(\ln ^2\left(\frac{1-w}{1+w}\right)+\pi^2\right)
 \\
 -4 w^2
      \left(447 w^6-430 w^4-88 w^2+48\right)
      \biggr)
\\      
      -384 \left(55 w^6-97 w^4+28  w^2+6\right) w^3
	 I_2\left(\sqrt{\frac{1-w}{1+w}}\right)
\\
+96 \left(55  w^4-110 w^2+51\right) w^6
	    I_1\left(\sqrt{\frac{1-w}{1+w}}\right)
\\
 +12 \left(-227 w^8+77 w^6+224 w^4+52 w^2-120\right) w 
 \times 
\\
\ln
	          \left(\frac{1-w}{1+w}\right)
+\sqrt{1-w^2} 
\biggl(
\pi \biggl(2048 w^6 \ln(2) 
\\
  +\frac{32}{3} \left(495 w^4-757 w^2+54\right)
		     w^4\biggr)
    -16 \pi ^2 
\\
\times
\left(165 w^8-220 w^6+21 w^4+30 w^2-36\right)
    \biggr)
\biggr] \,.
\end{multline}
Finally, the irreducible two-loop diagram, for the $2P$ level,
leads to the energy shift
\begin{multline}
f^{(2)}_{2P}(\beta)=\frac{1}{1152 w^6}
\biggl[
\biggl(1163 w^{10}-1966 w^8+75 w^6
\\
+184 w^4+64 w^2-192\biggr)
\left(\ln^2\left(\frac{1-w}{1+w}\right)+\pi ^2\right)
\\
-128 \left(105 w^6-191 w^4+52  w^2+6\right) w^3
      I_2\left(\sqrt{\frac{1-w}{1+w}}\right)
\\      
      +4 w^2
         \biggl(
	 24 \left(35 w^4-70 w^2+31\right) w^4
	    I_1\left(\sqrt{\frac{1-w}{1+w}}\right)
\\	    
    -937 w^6+978 w^4+88 w^2-48
    \biggr)
\\    
   -4 \left(517 w^8-299 w^6-400 w^4+44 w^2+120\right) w 
\times
\\
\ln\left(\frac{1-w}{1+w}\right)
    +\sqrt{1-w^2} \biggl(
    \pi  \biggl(
    2048 w^6 \ln (2)
\\    
    +\frac{32}{3} \left(315 w^4-529 w^2+18\right)
		     w^4\biggr)
\\
    -16 \pi ^2 \left(105 w^8-140 w^6+9 w^4-2 w^2-12\right)
    \biggr)
\biggr] \,.
\end{multline}
The difference of the $2P$ and $2S$ energy shifts
confirms Eq.~\eqref{muH} for muonic hydrogen.

\end{document}